\documentclass[11pt,graphicx,subfigure,axodraw]{article}
\usepackage{amsfonts}
\usepackage{amssymb}
\usepackage{amsmath,amsfonts,amssymb}
\usepackage{epstopdf}
\usepackage[section]{placeins}
\usepackage{hyperref}
\usepackage{booktabs}
\usepackage{cleveref}
\crefname{equation}{Eq.}{Eqs.}
\crefname{figure}{Fig.}{Figs.}
\crefname{table}{Table}{Tables}
\crefname{section}{Section}{Sections}

\usepackage[usenames,dvipsnames]{color}
\usepackage{graphicx}
\usepackage{subfigure}
\usepackage{float}
\def\rmuu{\gamma^{\mu}}
\def\rmud{\gamma_{\mu}}
\def\PL{{1-\gamma_5\over 2}}
\def\PR{{1+\gamma_5\over 2}}
\def\sinW2{\sin^2\theta_W}
\def\AEM{\alpha_{EM}}
\def\mul{M_{\tilde{u} L}^2}
\def\mur{M_{\tilde{u} R}^2}
\def\mdl{M_{\tilde{d} L}^2}
\def\mdr{M_{\tilde{d} R}^2}
\def\mz2{M_{z}^2}
\def\c2b{\cos 2\beta}
\def\au{A_u}
\def\ad{A_d}
\def\cob{\cot \beta}
\def\v#1{v_#1}
\def\tb{\tan\beta}
\def\epem{$e^+e^-$}
\def\KK{$K^0$-$\overline{K^0}$}
\def\wi{\omega_i}
\def\xj{\chi_j}
\def\Wmu{W_\mu}
\def\Wnu{W_\nu}
\def\m#1{{\tilde m}_#1}
\def\mH{m_H}
\def\mw#1{{\tilde m}_{\omega #1}}
\def\mx#1{{\tilde m}_{\chi^{0}_#1}}
\def\mc#1{{\tilde m}_{\chi^{+}_#1}}
\def\mwi{{\tilde m}_{\omega i}}
\def\mxi{{\tilde m}_{\chi^{0}_i}}
\def\mci{{\tilde m}_{\chi^{+}_i}}

\def\ch{{\tilde\chi^{+}_1}}
\def\c2{{\tilde\chi^{+}_2}}

\def\tt{{\tilde\theta}}

\def\tp{{\tilde\phi}}

\def\mz{M_z}
\def\sw{\sin\theta_W}
\def\cw{\cos\theta_W}
\def\cb{\cos\beta}
\def\sb{\sin\beta}
\def\rwi{r_{\omega i}}
\def\rxj{r_{\chi j}}
\def\rfp{r_f'}
\def\Kik{K_{ik}}
\def\Fq2{F_{2}(q^2)}
\def\f{\({\cal F}\)}
\def\d1{{\f(\tilde c;\tilde s;\tilde W)+ \f(\tilde c;\tilde \mu;\tilde W)}}
\def\tw{\tan\theta_W}
\def\sec2w{sec^2\theta_W}
\begin{document}
\baselineskip 18pt
\def\today{\ifcase\month\or
 January\or February\or March\or April\or May\or June\or
 July\or August\or September\or October\or November\or December\fi
 \space\number\day, \number\year}
\def\thebibliography#1{\section*{References\markboth
 {References}{References}}\list
 {[\arabic{enumi}]}{\settowidth\labelwidth{[#1]}
 \leftmargin\labelwidth
 \advance\leftmargin\labelsep
 \usecounter{enumi}}
 \def\newblock{\hskip .11em plus .33em minus .07em}
 \sloppy
 \sfcode`\.=1000\relax}
\let\endthebibliography=\endlist
\def\lsim{\ ^<\llap{$_\sim$}\ }
\def\gsim{\ ^>\llap{$_\sim$}\ }
\def\r2{\sqrt 2}
\def\beq{\begin{equation}}
\def\eeq{\end{equation}}
\def\beqn{\begin{eqnarray}}
\def\eeqn{\end{eqnarray}}
\def\rmuu{\gamma^{\mu}}
\def\rmud{\gamma_{\mu}}
\def\PL{{1-\gamma_5\over 2}}
\def\PR{{1+\gamma_5\over 2}}
\def\sinW2{\sin^2\theta_W}
\def\AEM{\alpha_{EM}}
\def\mul{M_{\tilde{u} L}^2}
\def\mur{M_{\tilde{u} R}^2}
\def\mdl{M_{\tilde{d} L}^2}
\def\mdr{M_{\tilde{d} R}^2}
\def\mz2{M_{z}^2}
\def\c2b{\cos 2\beta}
\def\au{A_u}
\def\ad{A_d}
\def\cob{\cot \beta}
\def\v#1{v_#1}
\def\tb{\tan\beta}
\def\epem{$e^+e^-$}
\def\KK{$K^0$-$\bar{K^0}$}
\def\wi{\omega_i}
\def\xj{\chi_j}
\def\Wmu{W_\mu}
\def\Wnu{W_\nu}
\def\m#1{{\tilde m}_#1}
\def\mH{m_H}
\def\mw#1{{\tilde m}_{\omega #1}}
\def\mx#1{{\tilde m}_{\chi^{0}_#1}}
\def\mc#1{{\tilde m}_{\chi^{+}_#1}}
\def\mwi{{\tilde m}_{\omega i}}
\def\mxi{{\tilde m}_{\chi^{0}_i}}
\def\mci{{\tilde m}_{\chi^{+}_i}}
\def\mz{M_z}
\def\sw{\sin\theta_W}
\def\cw{\cos\theta_W}
\def\cb{\cos\beta}
\def\sb{\sin\beta}
\def\rwi{r_{\omega i}}
\def\rxj{r_{\chi j}}
\def\rfp{r_f'}
\def\Kik{K_{ik}}
\def\Fq2{F_{2}(q^2)}
\def\f{\({\cal F}\)}
\def\d1{{\f(\tilde c;\tilde s;\tilde W)+ \f(\tilde c;\tilde \mu;\tilde W)}}
\def\tw{\tan\theta_W}
\def\sec2w{sec^2\theta_W}
\def\ch{{\tilde\chi^{+}_1}}
\def\c2{{\tilde\chi^{+}_2}}

\def\tt{{\tilde\theta}}

\def\tp{{\tilde\phi}}

\def\mz{M_z}
\def\sw{\sin\theta_W}
\def\cw{\cos\theta_W}
\def\cb{\cos\beta}
\def\sb{\sin\beta}
\def\rwi{r_{\omega i}}
\def\rxj{r_{\chi j}}
\def\rfp{r_f'}
\def\Kik{K_{ik}}
\def\Fq2{F_{2}(q^2)}
\def\f{\({\cal F}\)}
\def\d1{{\f(\tilde c;\tilde s;\tilde W)+ \f(\tilde c;\tilde \mu;\tilde W)}}

\def\b{$\cal{B}(\tau\to\mu \gamma)$~}

\def\ew{electroweak contribution to  the anomalous magnetic moment ~}

\def\tw{\tan\theta_W}
\def\sec2w{sec^2\theta_W}
\newcommand{\pn}[1]{{\color{red}{#1}}}

\begin{titlepage}
\begin{center}
{\large {\bf Probe of New Physics using Precision Measurement of the Electron Magnetic Moment}}\\


\vskip 0.5 true cm
Amin Aboubrahim$^{b}$\footnote{Email: amin.b@bau.edu.lb}, Tarek Ibrahim$^{a,b}$\footnote{Email:
t.ibrahim@bau.edu.lb},
  and Pran Nath$^{c}$\footnote{Emal: nath@neu.edu}
\vskip 0.5 true cm
\end{center}

\noindent
{$^{a}$ Department of  Physics, Faculty of Science,
University of Alexandria, Alexandria 21511, Egypt\footnote{Permanent address.} }\\
{$^{b}$ Department of Physics, Faculty of Sciences, Beirut Arab University,
Beirut 11 - 5020, Lebanon\footnote{Current address.}} \\
{$^{c}$ Department of Physics, Northeastern University, Boston, Massachusetts  02115-5000, USA} \\

\vskip 0.5 true cm

\centerline{\bf Abstract}

The anomalous  magnetic moment of the electron is    determined  experimentally   with an
accuracy of $2.8\times 10^{-13}$ and the uncertainty may decrease by an order of magnitude
in the future. While the current data is in excellent agreement with the standard model,
the  possible future improvement in the error in $\Delta a_e= a_e^{\text{exp}}- a_e^{\text{theory}}$
 has recently drawn interest in the electron anomalous magnetic moment
  as a possible probe
of new physics beyond the standard model.  In this work we give an analysis of such
physics in an
extension of the minimal supersymmetric standard model with
a vector multiplet. In the extended model the electroweak 
  contribution to the anomalous magnetic moment of the electron include loop diagrams involving in addition to the
exchange of W and Z,  the exchange of 
charginos, sneutrinos and mirror sneutrinos, and the exchange of neutralinos, sleptons and mirror sleptons. The analysis shows that a contribution to the electron magnetic moment
much larger than expected by $m_e^2/m_\mu^2$ scaling  of the deviation of the muon anomalous magnetic moment
over the standard model prediction, i.e., $\Delta a_\mu = 3 \times 10^{-9}$ as given by the
Brookhaven experiment, can be gotten within the MSSM extension.
Effects of CP violating phases in the extended MSSM model on the corrections to
the supersymmetric electroweak contributions  to $a_e$  are
also investigated. The analysis points to the  possibility of detection of
new physics effects with modest improvement on the error in $\Delta a_e= a_e^{\text{exp}} - a_e^{\text{theory}}$.

 \noindent
{\scriptsize
Keywords:{~Anomalous magnetic dipole moments, supersymmetry, vector lepton multiplets}\\
PACS numbers:~13.40Em, 12.60.-i, 14.60.Fg}

\medskip

\end{titlepage}
\section{Introduction \label{sec1}}
The anomalous  magnetic moment of the electron $a_e= (g-2)/2$ 
is one of most accurately determined quantities
experimentally. Thus the most recent determination of it gives the value ~\cite{Hanneke:2008tm}
\beqn
a_{e}^{\small \rm exp} = 115 \, 965 \, 218\, 07.3 \, (2.8) \times 10^{-13}.
\label{1.1}
\eeqn
In the standard model the contribution to the  magnetic moment  of the electron arises
from several sources so that (for a review see \cite{KM})

\beqn
a_e^{\rm SM} = a_e^{\rm qed} + a_e^{\rm EW} + a_e^{\rm had}
\label{1.2}
\eeqn
where $a_e^{\rm qed}$ involves purely QED  corrections and 
 includes one loop\cite{Schwinger:1948iu}, two loop \cite{qed2}, three loop \cite{qed3},
4 loop \cite{qed4,qed4a} and more recently five loop \cite{qed5a,qed5b,qed5c} contributions. 
Specifically in this work we will be using the results of  \cite{qed5b} which is an impressive
work giving the complete up to tenth -order QED contribution to $(g-2)_e$.
The analysis of \cite{qed5b} also improves the eighth-order contribution which includes
the mass-dependent contributions.
(In the context of precision analyses of $(g-2)_e$ and their sensitivity to higher terms in the 
ratio of the masses see also~\cite{qed4a,Kurz:2013exa}).
 $a_e^{\rm EW}$\cite{ckm} contains the
electroweak corrections involving the $W$ and $Z$ loops, and  $a_e^{\rm had}$ contains the
hadronic corrections ~\cite{Nomura:2012sb,Jegerlehner:2009ry}.
Now  in comparing the theory prediction with experiment one must use in the computation of the
theory prediction the value of $\alpha$
obtained in independent experiment rather than by equating $a_e^{\rm SM}(\alpha) = a_e^{\rm exp}$
~\cite{qed5b,Giudice:2012ms}.
Thus the analysis of  \cite{qed5b}
uses the value of $\alpha$ obtained from the measurement of $h/m_{\rm Rb}$
\cite{Bouchendira:2010es} 
combined with the accurately known Rydberg constant
and $m_{\rm Rb}/m_e$ 
 (for a review see~\cite{Mohr:2012tt}) which gives
 \beqn
\alpha^{-1}(\mathrm{^{87}Rb}) &\!\!\!=&
\!\!\! 1/137.035 \, 999 \, 049 \, (90)\,
\label{1.4}
\eeqn
Using Eq.(\ref{1.4}) the analysis of  \cite{qed5b} gives
\beqn
    a_e^{\scriptscriptstyle \rm SM} \!=\!
   115 \, 965 \, 218 \, 1.78 \, (6) (4) (3) (77) \times 10^{-12}\ , 
   \label{1.5}
\eeqn
where the numbers in the parentheses are as follows: (6) refers to the uncertainty in the four loop
QED coefficient, (4) refers to the uncertainty in the five loop QED co-efficient, (3) is the error
in the hadronic contribution, and (77) arises from the error in the determination of $\alpha$
using $\rm ^{87}Rb$ data. Combining the errors in quadratures  one finds  \cite{qed5b} that
 the uncertainty $\delta\Delta a_e$,  where $\Delta a_e =   (a_e^{\rm exp} - a_e^{\rm SM})$,
is given by
\beqn
\delta \Delta a_e =
8.2 \times 10^{-13}
\label{1.6}
\eeqn
Similar to the work of \cite{Giudice:2012ms} our motivation is to use Eq.(\ref{1.6}) to constrain new physics.
Specifically we look now at the implications of Eq.(\ref{1.6}) in view of the current status of the anomalous magnetic
moment of the muon.
 Thus the Brookhaven experiment indicates
a $\sim 3.5\sigma$ deviation from the standard model prediction, i.e., one has for $\Delta a_{\mu}$
the result~\cite{Hagiwara:2011af,Davier:2010nc}
\beqn
\Delta a_{\mu} = (287 \pm 80) \times 10^{-11}
\label{1.7}
\eeqn
Scaling the result of Eq.(\ref{1.7}) to the case of the electron by using the naive scaling factor of 
$m_e^2/m_\mu^2$ one gets a correction of size $(0.6 \pm 0.2)\times 10^{-13}$ which is an order of magnitude
smaller than the result of Eq.(\ref{1.6}). The above discussion indicates that if there are new physics
effects larger than those given by naive scaling, they
would be susceptible to discovery with modest improvements in the error $\delta \Delta a_e$.\\

In this work we carry out a detailed analysis of corrections to the anomalous
magnetic moment of the electron in extensions of MSSM with a vector multiplet (For a non-supersymmetric
analysis see also~\cite{Kannike:2011ng,Giudice:2012ms}).
 The analysis
will include contributions from the W and Z boson  loops, as well as corrections from
charginos, sneutrinos and mirror sneutrinos, from  neutralinos and sleptons and mirror sleptons.
It will be shown that the new physics
 corrections here
 can be far in excess of those implied by scaling and are
of a size that could be detectable in modest improvement in $\delta \Delta a_e$.
We also investigate the dependence of the anomalous magnetic moment of the electron on
CP phases arising from the supersymmetric contributions from the exchange of the vectorlike
multiplet. 
 In previous analyses within MSSM
 the supersymmetric correction to the
anomalous magnetic moment of the muon was found to be sensitive to CP phases in a
significant way~\cite{Ibrahim:1999aj} and we could have similar large CP dependent effects
for $\Delta a_e({\rm EW})$ 
in the analysis based on the MSSM extension. \\

The outline of the rest of the paper is as follows: In \cref{sec2} we discuss the MSSM extension with a
 vectorlike multiplet. Here we define the notation labeling the extra vectorlike particles, give their
 transformation properties under the SM gauge group and give the superpotential for the extended
 model. The D terms and the soft terms allowed in the model are discussed.
  In  \cref{sec3}
the interactions of leptons-sneutrinos (mirror sneutrinos)-charginos in the mass diagonal basis are
given. These interactions are used in the  computation of the left diagram of \cref{fig1}.
 In  \cref{sec4}
the interactions of leptons-sleptons (mirror sleptons)-neutralinos in the mass diagonal basis are
given. These interactions are used in the  computation of the right diagram of \cref{fig1}.
In \cref{sec5} the interactions of the  W and Z bosons that are needed in the computation of the
loop diagram of \cref{fig2} are discussed.
In \cref{sec6} an analytic  analysis  is given of the neutralino exchange contributions using the
interactions of \cref{sec3} and chargino exchange contribution using the interaction of \cref{sec4}.
Here an analytic analysis is also given of the exchange contributions of the W and Z bosons using
the interactions of \cref{fig2}.
 A detailed numerical analysis is given in \cref{sec7} for the electroweak 
 contribution to the electron anomalous magnetic moment in the model.  Here  the dependence of \ew of the electron  
 on supersymmetric CP phases is also investigated. It is shown that modest improvements
 in the current errors in $\Delta a_e$ can begin to probe the possible new physics contributions.
Further,  a relative comparison of 
 the electroweak contributions to the anomalous magnetic moments of $e,\mu,\tau$ is also given. 
  Conclusions are given in \cref{sec8}.
 Further details on the mass matrices for the sleptons and mirror sleptons are given in \cref{sec9}.

\section{MSSM Extension with a vector leptonic multiplet\label{sec2}}

Vector like multiplets arise in a variety of unified models~\cite{guts} some of which could be low lying.
They have been used recently in a variety of analyses
\cite{Babu:2008ge,Liu:2009cc,Martin:2009bg,Aboubrahim:2013yfa,Aboubrahim:2013gfa,Ibrahim:2008gg,Ibrahim:2010va,Ibrahim:2011im,Ibrahim:2010hv,Ibrahim:2012ds,Ibrahim:2009uv}.
In the analysis below we will assume an extended MSSM with just one leptonic vector mulitplet.
The addition of a vector multiplet keeps the model anomaly free. Before proceeding further we
define the notation and give a very brief description of the extended model and  a more detailed
description can be found in the previous works mentioned above. Thus the extended MSSM has
contains a vectorlike multiplet with the transformations   under $SU(3)_C\times SU(2)_L\times U(1)_Y$
as given below

 \begin{align}
\psi_{iL}\equiv
 \left(\begin{matrix} \nu_{i L}\cr
 ~{l}_{iL}  \end{matrix} \right) &&   l^c_{iL}&& \nu^c_{i L}\nonumber\\
 (1,2,- \frac{1}{2})&&(1,1,1)&&(1,1,0)\ .
\label{2}
\end{align}
where the last entry on the right hand side column  is the value of the hypercharge
 $Y$ defined so that $Q=T_3+ Y$.  These leptons have $V-A$ interactions.
We can now add a vectorlike multiplet where we have a fourth family of leptons with $V-A$ interactions
whose transformations can be gotten from Eq.(\ref{2}) by letting i run from 1-4.
A vectorlike lepton multiplet also has  mirrors and so we consider these mirror
leptons which have $V+A$ interactions. Its quantum numbers are given by

\begin{align}
\chi^c\equiv
 \left(\begin{matrix} E_{ L}^c \cr
 N_L^c\end{matrix}\right)&&  E_L  && N_L\nonumber\\
(1,2,\frac{1}{2}) && (1,1,-1) && (1,1,0).
\label{3}
\end{align}

Interesting new physics arises when we allow mixings of the vectorlike generation with
the three ordinary generations.  Thus the  superpotential of the model allowing for the mixings
among the three ordinary generations and the vectorlike generation is given by

\begin{align}
W&= -\mu \epsilon_{ij} \hat H_1^i \hat H_2^j+\epsilon_{ij}  [f_{1}  \hat H_1^{i} \hat \psi_L ^{j}\hat \tau^c_L
 +f_{1}'  \hat H_2^{j} \hat \psi_L ^{i} \hat \nu^c_{\tau L}
+f_{2}  \hat H_1^{i} \hat \chi^c{^{j}}\hat N_{L}
 +f_{2}'  H_2^{j} \hat \chi^c{^{i}} \hat E_{ L} \nonumber \\
&+ h_{1}  H_1^{i} \hat\psi_{\mu L} ^{j}\hat\mu^c_L
 +h_{1}'  H_2^{j} \hat\psi_{\mu L} ^{i} \hat\nu^c_{\mu L}
+ h_{2}  H_1^{i} \hat\psi_{e L} ^{j}\hat e^c_L
 +h_{2}'  H_2^{j} \hat\psi_{e L} ^{i} \hat\nu^c_{e L}] \nonumber \\
&+ f_{3} \epsilon_{ij}  \hat\chi^c{^{i}}\hat\psi_L^{j}
 + f_{3}' \epsilon_{ij}  \hat\chi^c{^{i}}\hat\psi_{\mu L}^{j}
 + f_{4} \hat\tau^c_L \hat E_{ L}  +  f_{5} \hat\nu^c_{\tau L} \hat N_{L}
 + f_{4}' \hat\mu^c_L \hat E_{ L}  +  f_{5}' \hat\nu^c_{\mu L} \hat N_{L} \nonumber \\
&+ f_{3}'' \epsilon_{ij}  \hat\chi^c{^{i}}\hat\psi_{e L}^{j}
 + f_{4}'' \hat e^c_L \hat E_{ L}  +  f_{5}'' \hat\nu^c_{e L} \hat N_{L}\ ,
 \label{5}
\end{align}
where  $\hat ~$ implies superfields,  $\hat\psi_L$ stands for $\hat\psi_{3L}$, $\hat\psi_{\mu L}$ stands for $\hat\psi_{2L}$
and  $\hat\psi_{e L}$ stands for $\hat\psi_{1L}$.
The mass terms for the neutrinos, mirror neutrinos,  leptons and  mirror leptons arise from the term
\beq
{\cal{L}}=-\frac{1}{2}\frac{\partial ^2 W}{\partial{A_i}\partial{A_j}}\psi_ i \psi_ j+\text{H.c.}
\label{6}
\eeq
where $\psi$ and $A$ stand for generic two-component fermion and scalar fields.
After spontaneous breaking of the electroweak symmetry, ($\langle H_1^1 \rangle=v_1/\sqrt{2} $ and $\langle H_2^2\rangle=v_2/\sqrt{2}$),
we have the following set of mass terms written in the 4-component spinor notation
so that
\beq
-{\cal L}_m= \bar\xi_R^T (M_f) \xi_L +\bar\eta_R^T(M_{\ell}) \eta_L +\text{H.c.},
\eeq
where the basis vectors in which the mass matrix is written is given by
\begin{gather}
\bar\xi_R^T= \left(\begin{matrix}\bar \nu_{\tau R} & \bar N_R & \bar \nu_{\mu R}
&\bar \nu_{e R} \end{matrix}\right),\nonumber\\
\xi_L^T= \left(\begin{matrix} \nu_{\tau L} &  N_L &  \nu_{\mu L}
& \nu_{e L} \end{matrix}\right) \ ,\nonumber\\
\bar\eta_R^T= \left(\begin{matrix}\bar{\tau_ R} & \bar E_R & \bar{\mu_ R}
&\bar{e_ R} \end{matrix}\right),\nonumber\\
\eta_L^T= \left(\begin{matrix} {\tau_ L} &  E_L &  {\mu_ L}
& {e_ L} \end{matrix}\right) \ ,
\end{gather}
and the mass matrix $M_f$ is given by

\beqn
M_f=
 \left(\begin{matrix} f'_1 v_2/\sqrt{2} & f_5 & 0 & 0 \cr
 -f_3 & f_2 v_1/\sqrt{2} & -f_3' & -f_3'' \cr
0&f_5'&h_1' v_2/\sqrt{2} & 0 \cr
0 & f_5'' & 0 & h_2' v_2/\sqrt{2}\end{matrix} \right)\ .
\label{7}
\eeqn
We define the matrix element $(22)$ of the mass matrix as $m_N$ so that 
\beqn
m_N= f_2 v_1/\sqrt 2.
\eeqn 
The mass matrix is not hermitian and thus one needs bi-unitary transformations to diagonalize it.
We define the bi-unitary transformation so that

\beq
D^{\nu \dagger}_R (M_f) D^\nu_L=\text{diag}(m_{\psi_1},m_{\psi_2},m_{\psi_3}, m_{\psi_4} ).
\label{7a}
\eeq
Under the bi-unitary transformations the basis vectors transform so that
\beqn
 \left(\begin{matrix} \nu_{\tau_R}\cr
 N_{ R} \cr
\nu_{\mu_R} \cr
\nu_{e_R} \end{matrix}\right)=D^{\nu}_R \left(\begin{matrix} \psi_{1_R}\cr
 \psi_{2_R}  \cr
\psi_{3_R} \cr
\psi_{4_R}\end{matrix}\right), \  \
\left(\begin{matrix} \nu_{\tau_L}\cr
 N_{ L} \cr
\nu_{\mu_L} \cr
\nu_{e_L}\end{matrix} \right)=D^{\nu}_L \left(\begin{matrix} \psi_{1_L}\cr
 \psi_{2_L} \cr
\psi_{3_L} \cr
\psi_{4_L}\end{matrix}\right) \ .
\label{8}
\eeqn
{
In \cref{7a}
$\psi_1, \psi_2, \psi_3, \psi_4$ are the mass eigenstates for the neutrinos,
where in the limit of no mixing
we identify $\psi_1$ as the light tau neutrino, $\psi_2$ as the
heavier mass eigen state,  $\psi_3$ as the muon neutrino and $\psi_4$ as the electron neutrino.
A similar analysis goes to the lepton mass matrix $M_\ell$ where
\beqn
M_\ell=
 \left(\begin{matrix} f_1 v_1/\sqrt{2} & f_4 & 0 & 0 \cr
 f_3 & f'_2 v_2/\sqrt{2} & f_3' & f_3'' \cr
0&f_4'&h_1 v_1/\sqrt{2} & 0 \cr
0 & f_4'' & 0 & h_2 v_1/\sqrt{2}\end{matrix} \right)\ .
\label{7}
\eeqn
We introduce now the mass parameter $m_E$ defined by the (22) element of the mass matrix above so that
\beqn
m_E=  f_2' v_2/\sqrt 2.
\eeqn
  Next we  consider  the mixing of the charged sleptons and the charged mirror sleptons.
The mass squared  matrix of the slepton - mirror slepton comes from three sources:  the F term, the
D term of the potential and the soft susy breaking terms.
Using the  superpotential of \cref{5} the mass terms arising from it
after the breaking of  the electroweak symmetry are given by
the Lagrangian
\beq
{\cal L}= {\cal L}_F +{\cal L}_D + {\cal L}_{\rm soft}\ ,
\eeq
where   $ {\cal L}_F$ is deduced from \cref{5} and is given in the Appendix, while the ${\cal L}_D$ is given by
\begin{align}
-{\cal L}_D&=\frac{1}{2} m^2_Z \cos^2\theta_W \cos 2\beta \{\tilde \nu_{\tau L} \tilde \nu^*_{\tau L} -\tilde \tau_L \tilde \tau^*_L
+\tilde \nu_{\mu L} \tilde \nu^*_{\mu L} -\tilde \mu_L \tilde \mu^*_L
+\tilde \nu_{e L} \tilde \nu^*_{e L} -\tilde e_L \tilde e^*_L \nonumber \\
&+\tilde E_R \tilde E^*_R -\tilde N_R \tilde N^*_R\}
+\frac{1}{2} m^2_Z \sin^2\theta_W \cos 2\beta \{\tilde \nu_{\tau L} \tilde \nu^*_{\tau L}
 +\tilde \tau_L \tilde \tau^*_L
+\tilde \nu_{\mu L} \tilde \nu^*_{\mu L} +\tilde \mu_L \tilde \mu^*_L \nonumber \\
&+\tilde \nu_{e L} \tilde \nu^*_{e L} +\tilde e_L \tilde e^*_L
-\tilde E_R \tilde E^*_R -\tilde N_R \tilde N^*_R +2 \tilde E_L \tilde E^*_L -2 \tilde \tau_R \tilde \tau^*_R
-2 \tilde \mu_R \tilde \mu^*_R -2 \tilde e_R \tilde e^*_R
\}.
\label{12}
\end{align}
For ${\cal L}_{\rm soft}$ we assume the following form
\begin{align}
-{\cal L}_{\text{soft}}&=\tilde M^2_{\tau L} \tilde \psi^{i*}_{\tau L} \tilde \psi^i_{\tau L}
+\tilde M^2_{\chi} \tilde \chi^{ci*} \tilde \chi^{ci}
+\tilde M^2_{\mu L} \tilde \psi^{i*}_{\mu L} \tilde \psi^i_{\mu L}
+\tilde M^2_{e L} \tilde \psi^{i*}_{e L} \tilde \psi^i_{e L}
+\tilde M^2_{\nu_\tau} \tilde \nu^{c*}_{\tau L} \tilde \nu^c_{\tau L}
 +\tilde M^2_{\nu_\mu} \tilde \nu^{c*}_{\mu L} \tilde \nu^c_{\mu L} \nonumber \\
&+\tilde M^2_{\nu_e} \tilde \nu^{c*}_{e L} \tilde \nu^c_{e L}
+\tilde M^2_{\tau} \tilde \tau^{c*}_L \tilde \tau^c_L
+\tilde M^2_{\mu} \tilde \mu^{c*}_L \tilde \mu^c_L
+\tilde M^2_{e} \tilde e^{c*}_L \tilde e^c_L
+\tilde M^2_E \tilde E^*_L \tilde E_L
 + \tilde M^2_N \tilde N^*_L \tilde N_L \nonumber \\
&+\epsilon_{ij} \{f_1 A_{\tau} H^i_1 \tilde \psi^j_{\tau L} \tilde \tau^c_L
-f'_1 A_{\nu_\tau} H^i_2 \tilde \psi ^j_{\tau L} \tilde \nu^c_{\tau L}
+h_1 A_{\mu} H^i_1 \tilde \psi^j_{\mu L} \tilde \mu^c_L
-h'_1 A_{\nu_\mu} H^i_2 \tilde \psi ^j_{\mu L} \tilde \nu^c_{\mu L} \nonumber \\
&+h_2 A_{e} H^i_1 \tilde \psi^j_{e L} \tilde e^c_L
-h'_2 A_{\nu_e} H^i_2 \tilde \psi ^j_{e L} \tilde \nu^c_{e L}
+f_2 A_N H^i_1 \tilde \chi^{cj} \tilde N_L
-f'_2 A_E H^i_2 \tilde \chi^{cj} \tilde E_L +\text{H.c.}\}\ .
\label{13}
\end{align}

\section{Interactions of leptons, scalar neutrinos and charginos\label{sec3}}
 In this section we discuss the  interactions in the mass diagonal basis involving charged leptons,
 sneutrinos and charginos.  Thus we have
\begin{align}
-{\cal L}_{\tau-\tilde{\nu}-\chi^{-}} &= \sum_{i=1}^{2}\sum_{j=1}^{8}\bar{\tau}_{\alpha}(C_{\alpha ij}^{L}P_{L}+C_{\alpha ij}^{R}P_{R})\tilde{\chi}^{ci}\tilde{\nu}_{j}+\text{H.c.},
\end{align}
such that
\begin{align}
\begin{split}
C_{\alpha ij}^{L}=&g(-\kappa_{\tau}U^{*}_{i2}D^{\tau*}_{R1\alpha} \tilde{D}^{\nu}_{1j} -\kappa_{\mu}U^{*}_{i2}D^{\tau*}_{R3\alpha}\tilde{D}^{\nu}_{5j}-
\kappa_{e}U^{*}_{i2}D^{\tau*}_{R4\alpha}\tilde{D}^{\nu}_{7j}+U^{*}_{i1}D^{\tau*}_{R2\alpha}\tilde{D}^{\nu}_{4j}-
\kappa_{N}U^{*}_{i2}D^{\tau*}_{R2\alpha}\tilde{D}^{\nu}_{2j})
\end{split} \\~\nonumber\\
\begin{split}
C_{\alpha ij}^{R}=&g(-\kappa_{\nu_{\tau}}V_{i2}D^{\tau*}_{L1\alpha}\tilde{D}^{\nu}_{3j}-\kappa_{\nu_{\mu}}V_{i2}D^{\tau*}_{L3\alpha}\tilde{D}^{\nu}_{6j}-
\kappa_{\nu_{e}}V_{i2}D^{\tau*}_{L4\alpha}\tilde{D}^{\nu}_{8j}+V_{i1}D^{\tau*}_{L1\alpha}\tilde{D}^{\nu}_{1j}+V_{i1}D^{\tau*}_{L3\alpha}\tilde{D}^{\nu}_{5j}\\
&+V_{i1}D^{\tau*}_{L4\alpha}\tilde{D}^{\nu}_{7j}-\kappa_{E}V_{i2}D^{\tau*}_{L2\alpha}\tilde{D}^{\nu}_{4j}),
\end{split}
\end{align}
with
\begin{align}
(\kappa_{N},\kappa_{\tau},\kappa_{\mu},\kappa_{e})&=\frac{(m_{N},m_{\tau},m_{\mu},m_{e})}{\sqrt{2}m_{W}\cos\beta} , \\~\nonumber\\
(\kappa_{E},\kappa_{\nu_{\tau}},\kappa_{\nu_{\mu}},\kappa_{\nu_{e}})&=\frac{(m_{E},m_{\nu_{\tau}},m_{\nu_{\mu}},m_{\nu_{e}})}{\sqrt{2}m_{W}\sin\beta} .
\end{align}

\section{Interactions of leptons, sleptons and neutralinos \label{sec4}}
 In this section we discuss the  interactions in the mass diagonal basis involving charged leptons,
 sleptons and neutralinos.  Thus we have

\begin{align}
-{\cal L}_{\tau-\tilde{\tau}-\chi^{0}} &= \sum_{i=1}^{4}\sum_{j=1}^{8}\bar{\tau}_{\alpha}(C_{\alpha ij}^{'L}P_{L}+C_{\alpha ij}^{'R}P_{R})\tilde{\chi}^{0}_{i}\tilde{\tau}_{j}+\text{H.c.},
\end{align}
such that
\begin{align}
C_{\alpha ij}^{'L}=&\sqrt{2}(\alpha_{\tau i}D^{\tau *}_{R1\alpha}\tilde{D}^{\tau}_{1j}-\delta_{E i}D^{\tau *}_{R2\alpha}\tilde{D}^{\tau}_{2j}-
\gamma_{\tau i}D^{\tau *}_{R1\alpha}\tilde{D}^{\tau}_{3j}+\beta_{E i}D^{\tau *}_{R2\alpha}\tilde{D}^{\tau}_{4j}
+\alpha_{\mu i}D^{\tau *}_{R3\alpha}\tilde{D}^{\tau}_{5j}-\gamma_{\mu i}D^{\tau *}_{R3\alpha}\tilde{D}^{\tau}_{6j} \nonumber\\
&+\alpha_{e i}D^{\tau *}_{R4\alpha}\tilde{D}^{\tau}_{7j}-\gamma_{e i}D^{\tau *}_{R4\alpha}\tilde{D}^{\tau}_{8j})
\end{align}
\begin{align}
C_{\alpha ij}^{'R}=&\sqrt{2}(\beta_{\tau i}D^{\tau *}_{L1\alpha}\tilde{D}^{\tau}_{1j}-\gamma_{E i}D^{\tau *}_{L2\alpha}\tilde{D}^{\tau}_{2j}-
\delta_{\tau i}D^{\tau *}_{L1\alpha}\tilde{D}^{\tau}_{3j}+\alpha_{E i}D^{\tau *}_{L2\alpha}\tilde{D}^{\tau}_{4j}
+\beta_{\mu i}D^{\tau *}_{L3\alpha}\tilde{D}^{\tau}_{5j}-\delta_{\mu i}D^{\tau *}_{L3\alpha}\tilde{D}^{\tau}_{6j}      \nonumber\\
&+\beta_{e i}D^{\tau *}_{L4\alpha}\tilde{D}^{\tau}_{7j}-\delta_{e i}D^{\tau *}_{L4\alpha}\tilde{D}^{\tau}_{8j}),
\end{align}
where

\begin{align}
\alpha_{E i}&=\frac{gm_{E}X^{*}_{4i}}{2m_{W}\sin\beta} \ ;  && \beta_{E i}=eX'_{1i}+\frac{g}{\cos\theta_{W}}X'_{2i}\left(\frac{1}{2}-\sin^{2}\theta_{W}\right) \\
\gamma_{E i}&=eX^{'*}_{1i}-\frac{g\sin^{2}\theta_{W}}{\cos\theta_{W}}X^{'*}_{2i} \  ;  && \delta_{E i}=-\frac{gm_{E}X_{4i}}{2m_{W}\sin\beta}
\end{align}

and
\begin{align}
\alpha_{\tau i}&=\frac{gm_{\tau}X_{3i}}{2m_{W}\cos\beta} \ ;  && \alpha_{\mu i}=\frac{gm_{\mu}X_{3i}}{2m_{W}\cos\beta} \ ; && \alpha_{e i}=\frac{gm_{e}X_{3i}}{2m_{W}\cos\beta}  \\
\delta_{\tau i}&=-\frac{gm_{\tau}X^{*}_{3i}}{2m_{W}\cos\beta} \ ; && \delta_{\mu i}=-\frac{gm_{\mu}X^{*}_{3i}}{2m_{W}\cos\beta} \ ; && \delta_{e i}=-\frac{gm_{e}X^{*}_{3i}}{2m_{W}\cos\beta}
\end{align}
{and where }

\begin{align}
\beta_{\tau i}=\beta_{\mu i}=\beta_{e i}&=-eX^{'*}_{1i}+\frac{g}{\cos\theta_{W}}X^{'*}_{2i}\left(-\frac{1}{2}+\sin^{2}\theta_{W}\right)  \\
\gamma_{\tau i}=\gamma_{\mu i}=\gamma_{e i}&=-eX'_{1i}+\frac{g\sin^{2}\theta_{W}}{\cos\theta_{W}}X'_{2i}
\end{align}
Here $X'$ are defined by

\begin{align}
X'_{1i}&=X_{1i}\cos\theta_{W}+X_{2i}\sin\theta_{W}  \\
X'_{2i}&=-X_ {1i}\sin\theta_{W}+X_{2i}\cos\theta_{W}
\end{align}
where $X$ diagonalizes the neutralino mass matrix, i.e.,

\beqn
X^{T}M_{\chi^{0}}X=\text{diag}(m_{\chi^{0}_{1}},m_{\chi^{0}_{2}},m_{\chi^{0}_{3}},m_{\chi^{0}_{4}}).
\eeqn

\section{Interaction of leptons and mirrors with W and Z bosons\label{sec5}}
In addition to the computation of the supersymmetric loop diagrams, we compute the contributions
arising from the exchange of the W and Z bosons and the leptons and the mirror leptons in the
loops. The relevant interactions needed are given below. For the W boson exchange the
interactions that enter are given by

\begin{align}
-{\cal L}_{\tau W\psi} &= W^{\dagger}_{\rho}\sum_{i=1}^{4}\sum_{\alpha=1}^{4}\bar{\psi}_{i}\gamma^{\rho}[C_{L_{i\alpha}}^W P_L + C_{R_{i\alpha}}^W P_R]\tau_{\alpha}+\text{H.c.}
\end{align}

where
\beqn
C_{L_{i\alpha}}^W= \frac{g}{\sqrt{2}} [D^{\nu*}_{L1i}D^{\tau}_{L1\alpha}+
D^{\nu*}_{L3i}D^{\tau}_{L3\alpha}+D^{\nu*}_{L4i}D^{\tau}_{L4\alpha}]  \\
C_{R_{i\alpha}}^W= \frac{g}{\sqrt{2}}[D^{\nu*}_{R2i}D^{\tau}_{R2\alpha}]
\eeqn
For the Z boson exchange the interactions that enter are given by

\beqn
-{\cal L}_{\tau\tau Z} &= Z_{\rho}\sum_{\alpha=1}^{4}\sum_{\beta=1}^{4}\bar{\tau}_{\alpha}\gamma^{\rho}[C_{L_{\alpha \beta}}^Z P_L + C_{R_{\alpha \beta}}^Z P_R]\tau_{\beta}
\eeqn
 where
\beqn
C_{L_{\alpha \beta}}^Z=\frac{g}{\cos\theta_{W}} [x(D_{L\alpha 1}^{\tau\dag}D_{L1\beta}^{\tau}+D_{L\alpha 2}^{\tau\dag}D_{L2\beta}^{\tau}+D_{L\alpha 3}^{\tau\dag}D_{L3\beta}^{\tau}+D_{L\alpha 4}^{\tau\dag}D_{L4\beta}^{\tau})\nonumber\\
-\frac{1}{2}(D_{L\alpha 1}^{\tau\dag}D_{L1\beta}^{\tau}+D_{L\alpha 3}^{\tau\dag}D_{L3\beta}^{\tau}+D_{L\alpha 4}^{\tau\dag}D_{L4\beta}^{\tau})]
\eeqn
and
\beqn
C_{R_{\alpha \beta}}^Z=\frac{g}{\cos\theta_{W}} [x(D_{R\alpha 1}^{\tau\dag}D_{R1\beta}^{\tau}+D_{R\alpha 2}^{\tau\dag}D_{R2\beta}^{\tau}+D_{R\alpha 3}^{\tau\dag}D_{R3\beta}^{\tau}+D_{R\alpha 4}^{\tau\dag}D_{R4\beta}^{\tau})\nonumber\\
-\frac{1}{2}(D_{R\alpha 2}^{\tau\dag}
D_{R 2\beta }^{\tau}
 )]
\eeqn

where $x=\sin^{2}\theta_{W}$.

\section{An analytical computation of the anomalous magnetic moment\label{sec6}}

\begin{figure}[h]
\begin{center}
{\rotatebox{0}{\resizebox*{10cm}{!}{\includegraphics{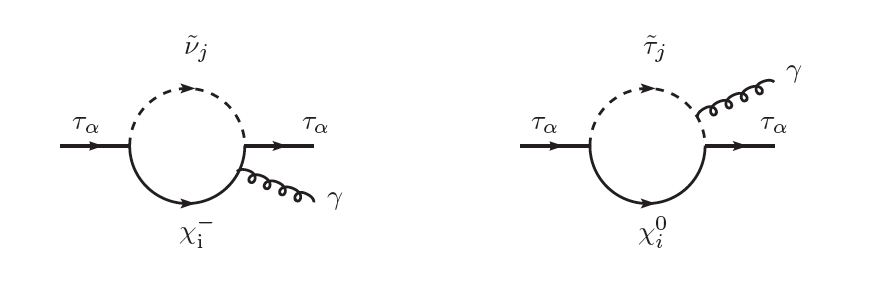}}\hglue5mm}}
\caption{The diagrams that contribute to the leptonic ($\tau_{\alpha}$)
magnetic dipole moment via
  exchange of charginos ($\chi_i^{-}$), sneutrinos and mirror sneutrinos  ($\tilde \nu_j$)   (left diagram) inside the loop and from the exchange
  of neutralinos ($\chi_i^0$)  sleptons   and mirror sleptons ($\tilde \tau_j$) (right diagram) inside the loop.} \label{fig1}
\end{center}
\end{figure}
Using the interactions given in \cref{sec3} the chargino contribution arises from the left diagram of
\cref{fig1}. It is given by

\begin{align}
a_{\alpha}^{\chi^{+}}&=-\sum_{i=1}^{2}\sum_{j=1}^{8}\frac{m_{\tau_{\alpha}}}{16\pi^{2}m_{\chi_{i}^{+}}}\text{Re}(C^{L}_{\alpha ij}C^{R*}_{\alpha ij})
F_{3}\left(\frac{m^{2}_{\tilde{\nu}_{j}}}{m^{2}_{\chi^{-}_{i}}}\right) \nonumber \\
&+\sum_{i=1}^{2}\sum_{j=1}^{8}\frac{m^{2}_{\tau_{\alpha}}}{96\pi^{2}m^{2}_{\chi_{i}^{+}}}\left[|C^{L}_{\alpha ij}|^{2}+|C^{R}_{\alpha ij}|^{2}\right]
F_{4}\left(\frac{m^{2}_{\tilde{\nu}_{j}}}{m^{2}_{\chi^{-}_{i}}}\right),
\end{align}
where the form factors $F_3$ and $F_4$ are given by

\begin{align}
F_{3}(x)&=\frac{1}{(x-1)^{3}}\left[3x^{2}-4x+1-2x^{2}\ln x \right]
\end{align}
and
\begin{align}
F_{4}(x)&=\frac{1}{(x-1)^{4}}\left[2x^{3}+3x^{2}-6x+1-6x^{2}\ln x \right]
\end{align}

Using the interactions given in \cref{sec4} the neutralino contribution arises from the right diagram of
\cref{fig1}. It is given by

\begin{align}
a_{\alpha}^{\chi^{0}}&=\sum_{i=1}^{4}\sum_{j=1}^{8}\frac{m_{\tau_{\alpha}}}{16\pi^{2}m_{\chi_{i}^{0}}}\text{Re}(C^{'L}_{\alpha ij}C^{'R*}_{\alpha ij})
F_{1}\left(\frac{m^{2}_{\tilde{\tau}_{j}}}{m^{2}_{\chi^{0}_{i}}}\right) \nonumber \\
&+\sum_{i=1}^{2}\sum_{j=1}^{8}\frac{m^{2}_{\tau_{\alpha}}}{96\pi^{2}m^{2}_{\chi_{i}^{0}}}\left[|C^{'L}_{\alpha ij}|^{2}+|C^{'R}_{\alpha ij}|^{2}\right]
F_{2}\left(\frac{m^{2}_{\tilde{\tau}_{j}}}{m^{2}_{\chi^{0}_{i}}}\right),
\end{align}
where the form factors are

\begin{align}
F_{1}(x)&=\frac{1}{(x-1)^{3}}\left[1-x^{2}+2x\ln x \right]
\end{align}
and
\begin{align}
F_{2}(x)&=\frac{1}{(x-1)^{4}}\left[-x^{3}+6x^{2}-3x-2-6x\ln x \right]
\end{align}
The anomalous magnetic moments are known to exhibit a sharp dependence on the
CP phases~\cite{Ibrahim:1999aj,Ibrahim:2007fb}. The dependence of $a_e$
on CP phases will
be exhibited in the numerical analysis to follow.

 The contributions to the lepton magnetic moment from the W and Z exchange arise from the
 diagrams of \cref{fig2}.  Using the interactions given in \cref{sec5} the contribution arising from
 the W exchange diagram (the left diagram of \cref{fig2}) is given by

\begin{figure}[h]
\begin{center}
{\rotatebox{0}{\resizebox*{10cm}{!}{\includegraphics{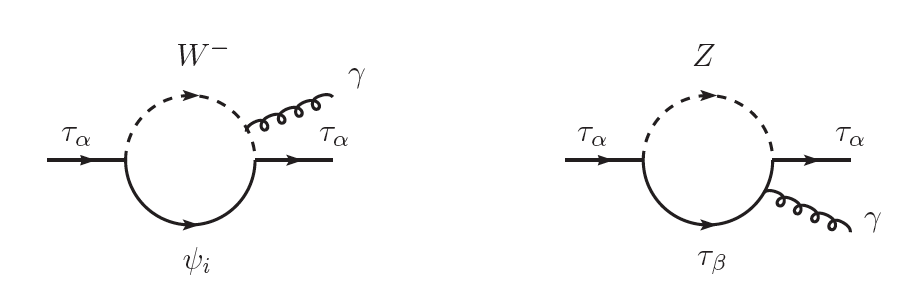}}\hglue5mm}}
\caption{ The W loop  (the left diagram) involving the exchange of sequential and vectorlike neutrinos $\psi_i$
and the Z loop (the right diagram) involving the exchange of sequential and vectorlike charged leptons $\tau_{\beta}$
that contribute to the magnetic moment of the charged lepton $\tau_{\alpha}$.}
\label{fig2}
\end{center}
\end{figure}

\beqn
a^W_{\tau_{\alpha}} = \frac{m^2_{\tau_{\alpha}}}{16 \pi^2 m^2_W} \sum_{i=1}^{4} [|C^W_{Li\alpha}|^2
+|C^W_{Ri\alpha}|^2] F_W \left(\frac{m^2_{\psi_i}}{m^2_W}\right) + \frac{m_{\psi_i}}{m_{\tau_{\alpha}}} \text{Re}(C^W_{L i \alpha}C^{W*}_{R i \alpha}) G_W \left(\frac{m^2_{\psi_i}}{m^2_W}\right),
\eeqn
where the form factors are given by
\begin{align}
F_{W}(x)&=\frac{1}{6(x-1)^{4}}\left[4 x^4- 49x^{3}+18 x^3 \ln x+78x^{2}-43 x +10 \right]
\end{align}
and
\begin{align}
G_{W}(x)&=\frac{1}{(x-1)^{3}}\left[4 -15 x+12 x^2 - x^3-6 x^2 \ln x \right]
\end{align}
Using the interactions given in \cref{sec5} the contribution arising from
 the Z exchange diagram (the right diagram of \cref{fig2}) is given by

\beqn
a^Z_{\tau_{\alpha}} = \frac{m^2_{\tau_{\alpha}}}{32 \pi^2 m^2_Z} \sum_{\beta=1}^{4} [|C^Z_{L \beta \alpha}|^2
+|C^Z_{R\beta \alpha}|^2] F_Z \left(\frac{m^2_{\tau_{\beta}}}{m^2_Z}\right) + \frac{m_{\tau_{\beta}}}{m_{\tau_{\alpha}}} \text{Re}(C^Z_{L \beta \alpha}C^{Z*}_{R \beta \alpha}) G_Z \left(\frac{m^2_{\tau_{\beta}}}{m^2_Z}\right),
\eeqn
where
\begin{align}
F_{Z}(x)&=\frac{1}{3(x-1)^{4}}\left[-5 x^4+14x^{3}-39 x^2+18 x^2 \ln x+38 x -8 \right]
\end{align}
and
\begin{align}
G_{Z}(x)&=\frac{2}{(x-1)^{3}}\left[x^3 + 3 x-6 x \ln x-4 \right].
\end{align}

We now show that the standard model result~\cite{fuji} 
can be gotten  in the  limit when the off diagonal elements in the neutrino and lepton mass matrices are set to zero. The W boson contribution is obtained in this case where the couplings are
$
C^W_{L i\alpha}= \dfrac{g}{\sqrt 2}$ for $ i=\alpha$ and zero othewise and
$C^W_{R i\alpha}=0$. In this limit, the form factor $F_W(0)= \frac{5}{3}$ and one gets
\beq
a^W_{\tau_{\alpha}}=\frac{5 g^2 m^2_{\tau_{\alpha}}}{96 \pi^2 m^2_W}
\eeq
Using the relation that $G_F = \dfrac{\pi \alpha_{em}}{\sqrt 2 m^2_W \sin^2\theta_W}$, one gets the
well known W boson contribution to the lepton $\tau_{\alpha}$
\beq
a^W_{\tau_{\alpha}}=\frac{5 G_F m^2_{\tau_{\alpha}}}{12\sqrt 2 \pi^2}
\label{aWSM}
\eeq
where $\alpha=3$ for the case of muon and $\alpha=4$ for the case of electron.\\

To recover the  Z boson contribution in the standard model limit we set
\beqn
C^Z_{L \beta \alpha} = \frac{g}{\cos\theta_W}\left(x - \frac{1}{2}\right)\ , \nonumber\\
C^Z_{R \beta \alpha} = \frac{g}{\cos \theta_W} x
\eeqn
for the case of $\alpha=\beta$ and are set to zero  otherwise. The form factors in this limit are given by $F_Z(0)=-\frac{8}{3}$ and $G_Z(0)= 8$. In this case one finds for the Z contribution, the well known Standard Model result

\beq
a^Z_{\tau_{\alpha}}=\frac{ G_F m^2_{\tau_{\alpha}}}{2\sqrt 2 \pi^2 } \left[-\frac{5}{12}+\frac{4}{3}\left(\sin^2\theta_W-\frac{1}{4}\right)^2\right].
\label{aZSM}
\eeq
\section{Numerical analysis and results\label{sec7}}

In this section we present a detailed numerical analysis of the effect of the extra vectorlike generation 
on the magnetic moment of the electron. We will also study the effects of CP phases on the 
electron magnetic moment. The analysis is done under the Brookhaven constraint on the anomalous
magnetic moment of the muon, i.e.,  the constraint of  Eq.(\ref{1.7}).
 As evident from the discussion of Section 2, the analysis is carried out in an MSSM extension with 
soft breaking parameters taken at the electroweak scale. Thus no renormalization group running of GUT scale
parameters  is needed. The parameters entering the analysis are summarized in the Appendix.
In \cref{table1}, we give a comparative analysis for the values of the electron anomalous magnetic moment for the case where no mixing occurs between generations and the case where such mixing takes place. 

\begin{table} \centering
\begin{tabular}{lcc}
\toprule\toprule
 & (i) Case of no mixing & (ii) Case of mixing \\ \cmidrule{2-3}
Chargino contribution & $1.52\times 10^{-14}$ & $5.76 \times 10^{-13}$  \\
Neutralino contribution & $7.03 \times 10^{-16}$ & $4.47 \times 10^{-16}$  \\
W boson contribution & $9.09 \times 10^{-14}$ & $1.02 \times 10^{-13}$  \\
Z boson contribution & $-4.59 \times 10^{-14}$ & $-3.89 \times 10^{-14}$  \\ \hline
$\Delta a_{e}$(EW) total  & $6.08 \times 10^{-14}$ & $6.39 \times 10^{-13}$  \\
\bottomrule \bottomrule
\end{tabular}
\caption{An exhibition the relative contributions to the electron magnetic dipole moment
arising from chargino exchange, neutralino exchange, W boson exchange and Z boson exchange
and their sum for the case when
 (i) there is no mixing among generations and for the case when (ii)  mixing occurs. The common parameters for the two cases are $m_{E}=250$, $m_{0}=m_{0}^{\tilde{\nu}}=650$, $|A_{0}|=520$, $|A_{0}^{\tilde{\nu}}|=650$, $|\mu|=102$, $|m_{1}|=600$, $|m_{2}|=680$, $\theta_{A_{0}}=1.2$, $\theta_{A_{0}^{\tilde{\nu}}}=2.8$, $\theta_{1}=2.5$, $\theta_{2}=1.5$, $\theta_{\mu}=0.5$, $m_{N}=212$ and $\tan\beta=15$. For case (i), the couplings $f_{3}=f'_{3}=f''_{3}=f_{4}=f'_{4}=f''_{4}=f_{5}=f'_{5}=f''_{5}=0$. For case (ii), the f couplings are non-zero and have the values $|f_{3}|=7\times 10^{-8}$, $|f'_{3}|=5 \times 10^{-8}$, $|f''_{3}|=8\times 10^{-9}$, $|f_{4}|=|f'_{4}|=10$, $|f''_{4}|=120$, $|f_{5}|=8.11\times 10^{-2}$, $|f'_{5}|=9.8 \times 10^{-2}$, $|f''_{5}|=4\times 10^{-2}$ and their phases are $\theta_{f_{3}}=0.3$, $\theta_{f'_{3}}=0.2$, $\theta_{f''_{3}}=0.6$, $\theta_{f_{4}}=1.4$, $\theta_{f'_{4}}=1.1$, $\theta_{f''_{4}}=0.5$, $\theta_{f_{5}}=1.9$, $\theta_{f'_{5}}=0.5$ and $\theta_{f''_{5}}=0.7$. All masses are in GeV and phases in rad.
A comparison of case (i) and case (ii) indicates a very significant increase for case (ii) overall.
The last row gives the total \ew which is sum of the four contributions in rows 1-4.  It is seen that the total
contribution for case (ii) is 10.5 times larger than for case (i).} \label{table1}
\end{table}

For  case (i) in \cref{table1}, the couplings $f_{3}$, $f'_{3}$, $f''_{3}$, $f_{4}$, $f'_{4}$, $f''_{4}$, $f_{5}$, $f'_{5}$ and $f''_{5}$ are all set to zero and this represents the case of no mixing between the generations. 
The upper two rows exhibit the chargino and neutralino contributions to $a_e$ while the next two rows 
give the standard model contribution arising from  $W$ and $Z$ exchanges given by Eqs.~(\ref{aWSM}) and~(\ref{aZSM}). The total $\Delta a_{e} ( {\rm EW})$ is shown in the bottom row and in this case it is $\sim 6\times 10^{-14}$. 
Case (ii) in \cref{table1} is when we include mixings between the vector generation and the sequential generations.
Here the $f$-couplings listed above assume non-zero values as shown in the table caption, indicating mixing between generations. The rows exhibit the  supersymmetric exchange contribution and the standard model contribution 
in same order as for case (i). The analysis shows that the total contribution from the electro-weak sector
increases by a factor of over 10 in the case when the mixing of the standard model generations with the vector multiplet is taken into
account.  The total electroweak correction in this case is $\Delta a_e ({\rm EW})=6.39\times 10^{-13}$ which lies 
just below the error corridor of Eq.(\ref{1.6}). The sparticles that enter the loops are neutralinos and sleptons, 
and charginos and sneutralinos. The current experimental  lower bound  on the neutralino mass  is $\sim 50$ GeV and
on the chargino and on  the slepton masses is $\sim 100$ GeV \cite{pdg}.  The analysis presented here respects these bounds.\\

\begin{figure}[H]
\begin{center}
{\rotatebox{0}{\resizebox*{10cm}{!}{\includegraphics{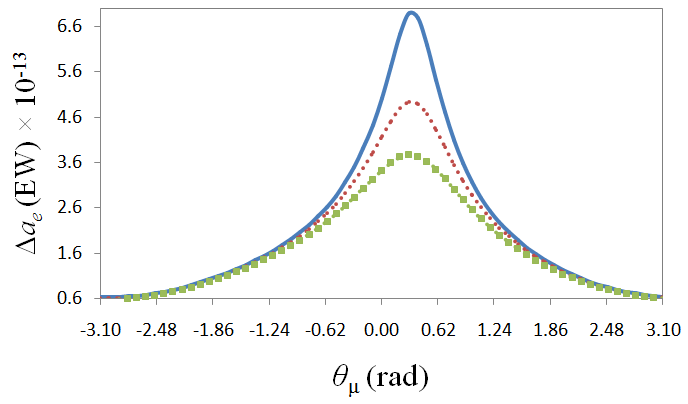}}\hglue5mm}}
\caption{A display of the electron  anomalous magnetic moment as a function of $\theta_{\mu}$, the phase of $\mu$, in the range $[-\pi,\pi]$. The three curves correspond to $m_{0}^{\tilde{\nu}}=650$, $|A_{0}^{\tilde{\nu}}|=650$, $|A_{0}|=520$ (solid curve), $m_{0}^{\tilde{\nu}}=660$, $|A_{0}^{\tilde{\nu}}|=655$, $|A_{0}|=530$ (dotted curve) and $m_{0}^{\tilde{\nu}}=675$, $|A_{0}^{\tilde{\nu}}|=660$, $|A_{0}|=540$ (square dotted curve). Other parameters have the values $m_{N}=212$, $m_{E}=250$, $m_{0}=650$, $|m_{1}|=600$, $|m_{2}|=240$, $|\mu|=104$, $\tan\beta=15$, $|f_{3}|=7\times 10^{-8}$, $|f'_{3}|=5 \times 10^{-8}$, $|f''_{3}|=8\times 10^{-9}$, $|f_{4}|=|f'_{4}|=10$, $|f''_{4}|=90$, $|f_{5}|=8.11\times 10^{-2}$, $|f'_{5}|=9.8 \times 10^{-2}$, $|f''_{5}|=4\times 10^{-2}$, $\theta_{A_{0}}=1.2$, $\theta_{A_{0}^{\tilde{\nu}}}=2.8$, $\theta_{1}=2.5$, $\theta_{2}=1.5$, $\theta_{f_{3}}=0.3$, $\theta_{f'_{3}}=0.2$, $\theta_{f''_{3}}=0.6$, $\theta_{f_{4}}=1.4$, $\theta_{f'_{4}}=1.1$, $\theta_{f''_{4}}=0.5$, $\theta_{f_{5}}=1.9$, $\theta_{f'_{5}}=0.5$ and $\theta_{f''_{5}}=0.7$. All masses are in GeV and phases in rad.}
\label{aemu}
\end{center}
\end{figure}

 It is known that the supersymmetric electroweak correction to the anomalous magnetic moment 
 is sensitive to CP phases. This was demonstrated for the case of the supersymmetric electroweak
 contributions to the anomalous magnetic moment of the muon in ~\cite{Ibrahim:1999aj}.
 Here we exhibit this sensitivity for the case of the electroweak contributions to the electron 
 anomalous magnetic moment. 
Thus 
Figure~\ref{aemu} displays the total \ew of the electron 
as a function of $\theta_{\mu}$ which is the phase of $\mu$ that appears in the chargino and neutralino mass matrices and in the slepton and sneutrino mass$^{2}$ matrices. Over the interval $[-\pi,\pi]$, the electroweak  correction to the 
 anomalous magnetic moment of the electron shows a pronounced peak for a value of $\theta_{\mu}=0.3$ rad. For the three sets of values considered, the peak values stretch from $\sim 3.7-6.9\times 10^{-13}$. 
 It should be noted that the variation in $a_{e}$ comes from the supersymmetric sector, and mainly from the chargino contribution. This is so because the neutralino contribution is relatively small, typically an order of magnitude
 smaller than the chargino exchange contribution. Because of this the variation of $a_e$ with  CP phases 
 is dominated by the chargino contributions. We note also that the $W$ and $Z$ contributions are not affected 
 by the phases. \\

\begin{figure}[H]
\begin{center}
{\rotatebox{0}{\resizebox*{10cm}{!}{\includegraphics{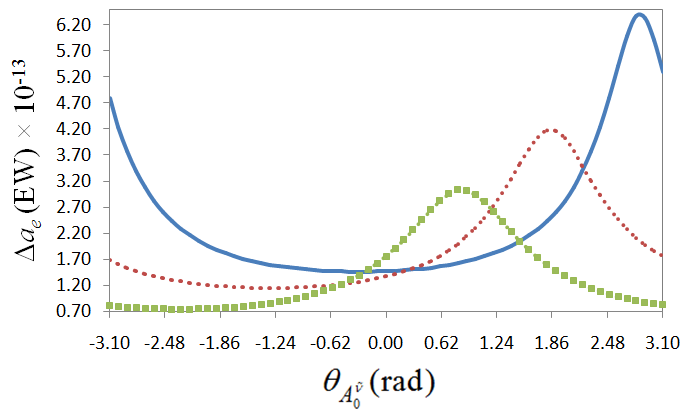}}\hglue5mm}}
\caption{A display of the \ew of the 
electron  as a function of $\theta_{A_{0}^{\tilde{\nu}}}$, the phase of $A_{0}^{\tilde{\nu}}$, in the range $[-\pi,\pi]$. The three curves correspond to (i) $m_{0}^{\tilde{\nu}}=650$, $\theta_{\mu}=0.3$ (solid curve), (ii) $m_{0}^{\tilde{\nu}}=660$, $\theta_{\mu}=1.3$ (dotted curve) and (iii) $m_{0}^{\tilde{\nu}}=670$, $\theta_{\mu}=2.3$ (square dotted curve). Other parameters have the values $m_{N}=212$, $m_{E}=250$, $m_{0}=650$, $|m_{1}|=600$, $|m_{2}|=240$, $|\mu|=103$, $\tan\beta=15$, $|A_{0}^{\tilde{\nu}}|=660$, $|A_{0}|=520$, $|f_{3}|=7\times 10^{-8}$, $|f'_{3}|=5 \times 10^{-8}$, $|f''_{3}|=8\times 10^{-9}$, $|f_{4}|=|f'_{4}|=10$, $|f''_{4}|=90$, $|f_{5}|=8.11\times 10^{-2}$, $|f'_{5}|=9.8 \times 10^{-2}$, $|f''_{5}|=4\times 10^{-2}$, $\theta_{A_{0}}=1.2$, $\theta_{1}=2.5$, $\theta_{2}=1.5$, $\theta_{f_{3}}=0.3$, $\theta_{f'_{3}}=0.2$, $\theta_{f''_{3}}=0.6$, $\theta_{f_{4}}=1.4$, $\theta_{f'_{4}}=1.1$, $\theta_{f''_{4}}=0.5$, $\theta_{f_{5}}=1.9$, $\theta_{f'_{5}}=0.5$ and $\theta_{f''_{5}}=0.7$. All masses are in GeV and phases in rad.}
\label{aeA0n}
\end{center}
\end{figure}

Figure~\ref{aeA0n} exhibits the variation of \ew of the electron
$\Delta a_{e}(\rm{EW})$ as a function of $\theta_{A_{0}^{\tilde{\nu}}}$, the phase of the trilinear coupling $A_{0}^{\tilde{\nu}}$, where in our analysis we have assumed that $A_{\nu_{\tau}}= A_{\nu_{\mu}}= A_{\nu_{e}}=A_{N}=A_{0}^{\tilde{\nu}}$ and $m_{0}^{\tilde{\nu}^{2}}=\tilde{M}_{N}^{2}=\tilde{M}_{\nu_{\tau}}^{2}=\tilde{M}_{\nu_{\mu}}^{2}=\tilde{M}_{\nu_{e}}^{2}$ in the sneutrino mass$^{2}$ matrix (see Appendix). Note that $m_{0}^{2}=\tilde{M_{\tau L}}^{2}=\tilde{M}_{E}^{2}=\tilde{M}_{\tau}^{2}=\tilde{M}_{\chi}^{2}=\tilde{M}_{\mu L}^{2}=\tilde{M}_{\mu}^{2}=\tilde{M}_{e L}^{2}=\tilde{M}_{e}^{2}$ and $A_{0}=A_{\tau}=A_{E}=A_{\mu}=A_{e}$ in the slepton mass$^{2}$ matrix (see Appendix). 
As can be seen from ~\cref{aeA0n} the variation is very substantial with $\Delta a_{e}(\rm{EW})$
varying in the range 
$\sim 7\times 10^{-14}-6\times 10^{-13}$ which is an order of magnitude variation. 
As for the case of \cref{aemu} the source of variation is the chargino exchange contribution once again.
This is so because the chargino exchange diagram contains the  sneutrino mass matrix in the loop 
which has a strong $A_{0}^{\tilde{\nu}}$ dependence.\\

\begin{figure}[H]
\begin{center}
\hfill
\subfigure[]
{\includegraphics[width=8cm]{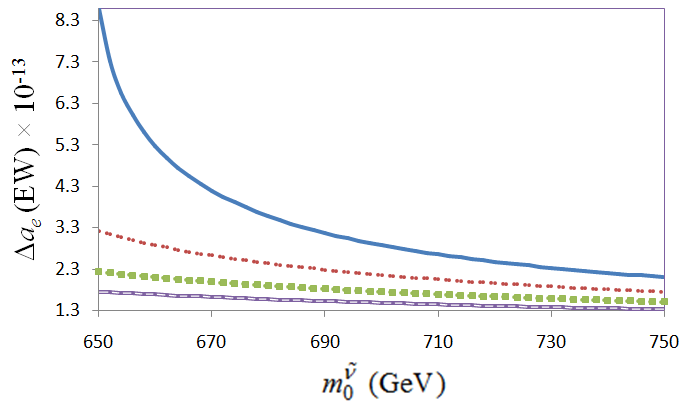}}
\hfill
\subfigure[]
{\includegraphics[width=8cm]{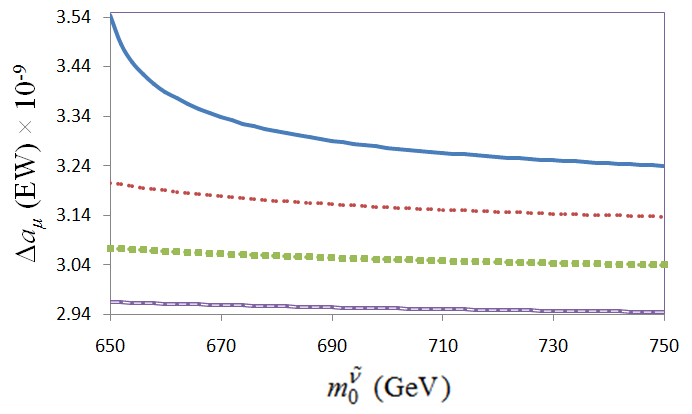}}
\hfill
\caption{A display of the \ew of $e,\mu$ 
as a function of $m_{0}^{\tilde{\nu}}$ in the range 650-750 GeV. Panel (a) gives the electron anomalous magnetic moment and panel (b) gives the muon anomalous magnetic moment. The curves correspond to $\tan\beta=13$ (lowermost curve), $\tan\beta=14$ (square dotted curve), $\tan\beta=15$ (dotted curve), and $\tan\beta=16$ (solid curve). Other parameters have the values $m_{N}=212$, $m_{E}=250$, $m_{0}=650$, $|m_{1}|=600$, $|m_{2}|=200$, $|\mu|=104$, $|A_{0}^{\tilde{\nu}}|=650$, $|A_{0}|=520$, $|f_{3}|=7\times 10^{-8}$, $|f'_{3}|=5 \times 10^{-8}$, $|f''_{3}|=8\times 10^{-9}$, $|f_{4}|=|f'_{4}|=10$, $|f''_{4}|=90$, $|f_{5}|=8.11\times 10^{-2}$, $|f'_{5}|=9.8 \times 10^{-2}$, $|f''_{5}|=4\times 10^{-2}$, $\theta_{A_{0}}=1.2$, $\theta_{A_{0}^{\tilde{\nu}}}=2.8$, $\theta_{1}=2.5$, $\theta_{2}=1.5$, $\theta_{\mu}=1.0$, $\theta_{f_{3}}=0.3$, $\theta_{f'_{3}}=0.2$, $\theta_{f''_{3}}=0.6$, $\theta_{f_{4}}=1.4$, $\theta_{f'_{4}}=1.1$, $\theta_{f''_{4}}=0.5$, $\theta_{f_{5}}=1.9$, $\theta_{f'_{5}}=0.5$ and $\theta_{f''_{5}}=0.7$. All masses are in GeV and phases in rad.} \label{am0n}
\end{center}
\end{figure}

Figure~\ref{am0n} exhibits the  variation of the \ew of the 
electron and of the muon  as a function of $m_{0}^{\tilde{\nu}}$ over the range $650-750$ GeV.
For parametric curves corresponding to $\tan\beta=13, 14, 15, 16$ (from bottom to top) are shown
for each of the panels. A comparison of panel (a) with panel (b) shows that  $a_e$ exhibits a much
larger sensitivity to $m_{0}^{\tilde{\nu}}$. 

\begin{figure}[H]
\begin{center}
\hfill
\subfigure[]
{\includegraphics[width=8cm]{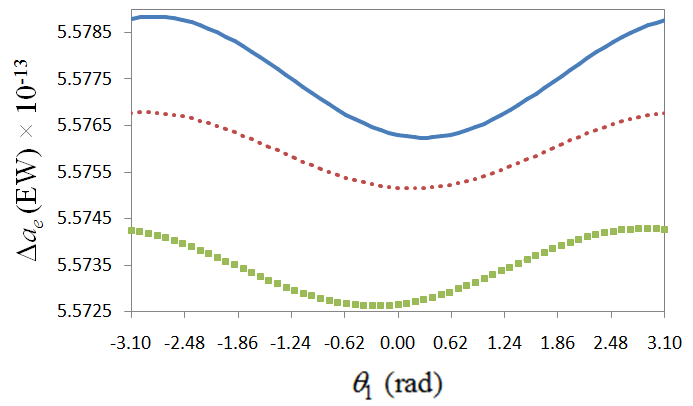}}
\hfill
\subfigure[]
{\includegraphics[width=8cm]{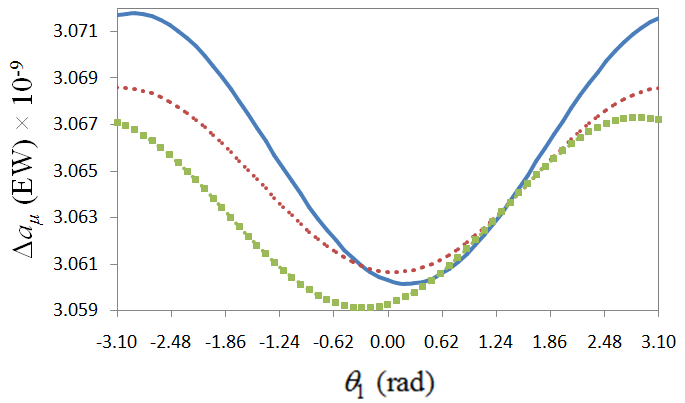}}
\hfill
\subfigure[]
{\includegraphics[width=8cm]{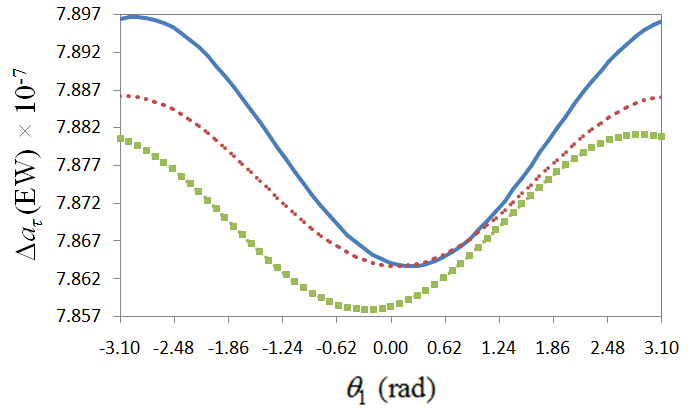}}
\caption{A display of the \ew of the $e,\mu,\tau$ 
 as a function of $\theta_{1}$, the phase of $m_{1}$, in the range $[-\pi,\pi]$. Panel (a) gives the 
 \ew of  the
 electron,  panel (b) gives it for the muon  and panel (c) gives it for the  tau. The curves correspond to (i) $|A_{0}|=520$, $\theta_{A_{0}}=1.2$, $|m_{2}|=350$ (solid curve), (ii) $|A_{0}|=330$, $\theta_{A_{0}}=1.9$, $|m_{2}|=351$ (dotted curve), and (iii) $|A_{0}|=130$, $\theta_{A_{0}}=2.6$, $|m_{2}|=352$ (lowermost curve). Other parameters have the values $m_{N}=212$, $m_{E}=250$, $m_{0}=m_{0}^{\tilde{\nu}}=650$, $|m_{1}|=600$, $|\mu|=104$, $|A_{0}^{\tilde{\nu}}|=640$, $\tan\beta=15$, $|f_{3}|=7\times 10^{-8}$, $|f'_{3}|=5 \times 10^{-8}$, $|f''_{3}|=8\times 10^{-9}$, $|f_{4}|=|f'_{4}|=10$, $|f''_{4}|=90$, $|f_{5}|=8.11\times 10^{-2}$, $|f'_{5}|=9.8 \times 10^{-2}$, $|f''_{5}|=4\times 10^{-2}$, $\theta_{2}=0.1$, $\theta_{A_{0}^{\tilde{\nu}}}=2.8$, $\theta_{\mu}=0.3$, $\theta_{f_{3}}=0.3$, $\theta_{f'_{3}}=0.2$, $\theta_{f''_{3}}=0.6$, $\theta_{f_{4}}=1.4$, $\theta_{f'_{4}}=1.1$, $\theta_{f''_{4}}=0.5$, $\theta_{f_{5}}=1.9$, $\theta_{f'_{5}}=0.5$ and $\theta_{f''_{5}}=0.7$. All masses are in GeV and phases in rad.} \label{am1}
\end{center}
\end{figure}

Figure~\ref{am1} exhibits the variation of the  \ew of the electron (panel (a)), of the muon (panel(b)) and
of the  tau (panel (c))  vs $\theta_{1}$, which is the phase of $m_{1}$, over the range $[-\pi,\pi]$. 
It seen that the variation is smooth in all cases as expected. However, the size of  variation in each case
is small as can be seen, for example, by comparing the range of variation in panel (a) in \cref{am1} with the range of 
variation in \cref{aemu}.  The reason for this smallness is easily understood. Thus the parameter
$m_1$ enters in the neutralino mass matrix and as discussed earlier the contribution from the
neutralino exchange diagram to the \ew of the electron is relatively small which explains the relative
smallness of the variation  of $\Delta a_e({\rm EW})$ with $\theta_1$. Similar results hold for the variation of 
$\Delta a_{\mu}({\rm EW})$ and $\Delta a_{\tau}(\rm{EW})$ with $\theta_1$. 
Regarding  $a_{\tau}$ we note that  the standard model predicts ~\cite{Eidelman:2007sb}
\beqn
a_{\tau}^{\rm SM} =  117 721 (5) \times 10^{-8}\ .
\eeqn
The current experimental result is~\cite{Abdallah:2003xd}
\beqn
a_{\tau}^{\rm EXP} = - 0.018 (17)\ , 
\eeqn
while the analysis of ~\cite{GonzalezSprinberg:2000mk} constraints the range of new physics so that
\beqn
 -0.007 < \Delta a_\tau^{\rm NP} <0.005\ , 
 \label{np}
\eeqn
where $\Delta a_\tau^{\rm NP}$ refers to the new physics contribution. 
Future experiments ~\cite{Lusiani:2011zz} in high luminosity B factories are likely to significantly 
improve the limits in Eq.(\ref{np}). 
However, it is unlikely that the improvements in the 
measurement of the tau anomalous magnetic moment 
at the level needed
to check on the contributions of  panel ( c) in \cref{am1} can be achieved in experiment in the very
near future. Thus, $a_e$ gives the best hope for the test of new physics.


 Next we investigate the limits on the parameter space arising from the constraints of
 Eq.(\ref{1.6}) and Eq.(\ref{1.7}).  Using these we impose the following upper limit constraints on
 new physics contributions 
 \beqn
 \label{conae}
 \Delta a_e^{\rm NP} \leq 8.2 \times 10^{-13}\\
 \Delta a_{\mu}^{\rm NP} \leq 2.87 \times 10^{-9}
 \label{conamu}
 \eeqn
where $\Delta a_e^{\rm NP}$ stands for the new physics contribution to $a_e$ and 
 $\Delta a_{\mu}^{\rm NP}$ stands for the new physics contribution to $a_\mu$. 
 In Fig.(\ref{e-mu-scatt}) we give an analysis of the allowed (shaded) and excluded (empty) regions
 under the constraints of Eqs.(\ref{conae}, \ref{conamu}).
Thus the left panel of Fig.(\ref{e-mu-scatt}) gives an analysis of the allowed parameter space in the
$m_0-m_2$ plane of the constraint on the new physics contribution to the anomalous magnetic moment of the electron  
given by Eqs.(\ref{conae})
while  the 
 right panel of Fig.(\ref{e-mu-scatt}) gives an analysis of the allowed parameter space in the
$m_0-m_2$ plane of the constraint on the new physics contribution to the anomalous magnetic moment of the 
muon given by Eqs.(\ref{conamu}). We see that a part of the parameter space is excluded because of the
constraints. The excluded regions are of course, sensitive to the other input parameters and different 
choices of those parameters would lead to modification of the allowed and the excluded regions. 
Next we exhibit the allowed and the excluded regions in the plane of CP phases.  
The analysis here is similar to the one in \cite{Ibrahim:2001ym} done for the muon anomaly.
Thus as noted earlier the anomalous magnetic moments are sensitive to CP phases and 
the constraint of Eq.(\ref{conae}) would have impact on the allowed regions of the CP phases.
This is exhibited in Fig.(\ref{escattp}) where the allowed (filled) and excluded (empty) regions 
under the constraint of Eq.(\ref{conae}) are exhibited in the plane of two phases: the phase of
the Higgs mixing parameter $\theta_\mu$ and the phase of the trilinear coupling $A_0^{\tilde \nu}$,
i.e., $\theta_{A_0^{\tilde \nu}}$. The analysis of Fig. (\ref{e-mu-scatt}) and Fig.(\ref{escattp}) indicates
that the constraints on new physics given by Eq.(\ref{conae})
and Eq.(\ref{conamu}) have significant impact on the available parameter space of the extended
MSSM model.

\begin{figure}[H]
\begin{center}
\hfill
\subfigure[]
{\includegraphics[width=8cm]{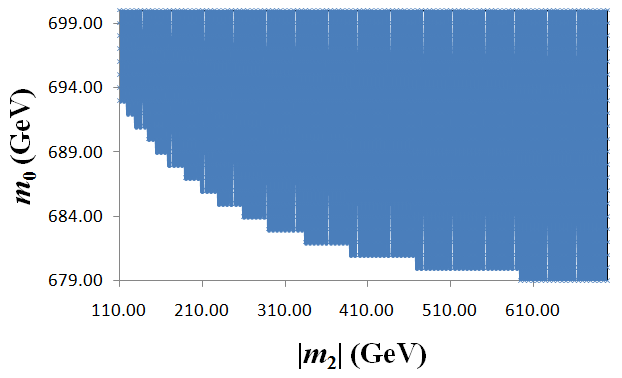}}
\hfill
\subfigure[]
{\includegraphics[width=8cm]{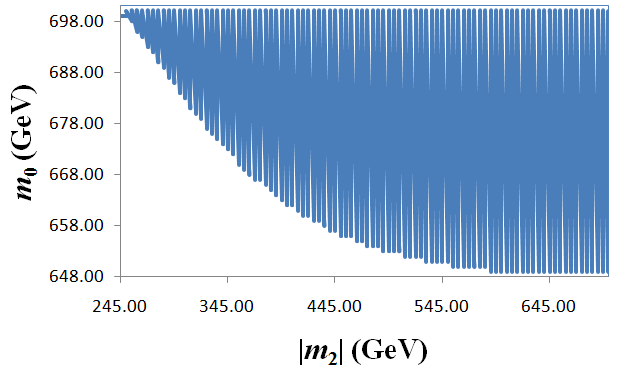}}
\hfill
\caption{
An exhibition of the allowed (shaded) and excluded (empty) regions in the $m_0-m_2$ plane under the anomalous magnetic moment constraints. 
Panel (a) gives a display of the allowed and forbidden values of $m_{0}$ and $|m_{2}|$ under the constraint on $a_{e}$ 
given by 
Eq.(\ref{conae})
and 
 panel (b) gives a display of the allowed and forbidden regions of $m_{0}$ and $|m_{2}|$ under the constraint on $a_{\mu}$ of 
Eq.(\ref{conamu}). 
 Other parameters have the values $\tan\beta=15$, $m_{N}=212$, $m_{E}=180$, $|m_{1}|=600$, $|\mu|=104$, $|A_{0}^{\tilde{\nu}}|=651$, $m_{0}^{\tilde{\nu}}=650$, $|A_{0}|=520$, $|f_{3}|=7\times 10^{-8}$, $|f'_{3}|=5 \times 10^{-8}$, $|f''_{3}|=8\times 10^{-9}$, $|f_{4}|=|f'_{4}|=10$, $|f''_{4}|=100$, $|f_{5}|=8.11\times 10^{-2}$, $|f'_{5}|=9.8 \times 10^{-2}$, $|f''_{5}|=4\times 10^{-2}$, $\theta_{A_{0}}=1.2$, $\theta_{A_{0}^{\tilde{\nu}}}=2.8$, $\theta_{1}=2.5$, $\theta_{2}=0$, $\theta_{\mu}=0.3$, $\theta_{f_{3}}=0.3$, $\theta_{f'_{3}}=0.2$, $\theta_{f''_{3}}=0.6$, $\theta_{f_{4}}=1.4$, $\theta_{f'_{4}}=1.1$, $\theta_{f''_{4}}=0.5$, $\theta_{f_{5}}=1.9$, $\theta_{f'_{5}}=0.5$ and $\theta_{f''_{5}}=0.7$. All masses are in GeV and phases in rad.} \label{e-mu-scatt}
\end{center}
\end{figure}

\begin{figure}[H]
\begin{center}
{\rotatebox{0}{\resizebox*{8cm}{!}{\includegraphics{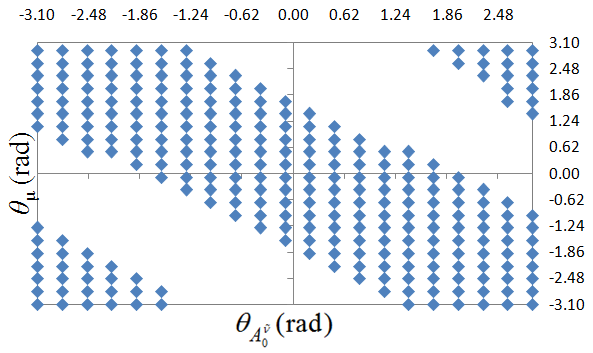}}\hglue5mm}}
\caption{
An exhibition of the allowed (shaded) and excluded (empty) regions in the
$\theta_{\mu}-\theta_{A_{0}^{\tilde{\nu}}}$ plane under the constraint on $a_{e}$ given by 
Eq.(\ref{conae}).
Other parameters have the values $\tan\beta=15$, $m_{N}=212$, $m_{E}=130$, $m_{0}=649$, $|m_{1}|=600$, $|m_{2}|=100$, $|\mu|=104$, $|A_{0}^{\tilde{\nu}}|=651$, $m_{0}^{\tilde{\nu}}=650$, $|A_{0}|=520$, $|f_{3}|=7\times 10^{-8}$, $|f'_{3}|=5 \times 10^{-8}$, $|f''_{3}|=8\times 10^{-9}$, $|f_{4}|=|f'_{4}|=10$, $|f''_{4}|=140$, $|f_{5}|=8.11\times 10^{-2}$, $|f'_{5}|=9.8 \times 10^{-2}$, $|f''_{5}|=4\times 10^{-2}$, $\theta_{A_{0}}=1.2$, $\theta_{1}=2.5$, $\theta_{2}=0$, $\theta_{f_{3}}=0.3$, $\theta_{f'_{3}}=0.2$, $\theta_{f''_{3}}=0.6$, $\theta_{f_{4}}=1.4$, $\theta_{f'_{4}}=1.1$, $\theta_{f''_{4}}=0.5$, $\theta_{f_{5}}=1.9$, $\theta_{f'_{5}}=0.5$ and $\theta_{f''_{5}}=0.7$. All masses are in GeV and phases in rad.}
\label{escattp}
\end{center}
\end{figure}

\section{Conclusion\label{sec8}}
The magnetic moment of the electron is one of the most precisely determined quantities in physics
with an error in $a_e^{exp}$  of $\delta a_e^{exp} = 2.8\times 10^{-13}$. The theory predictions for
$a_e$ have also been done with a high accuracy.  However the error in theory prediction is
  significantly larger than the
experimental error giving a total error in the difference in experiment minus theory of
$\delta \Delta a_e\simeq  8  \times 10^{-13}$.
This error is much larger than the new physics effects predicted by scaling if one extrapolates
the discrepancy between experiment and theory for the muon anomalous magnetic moment.
Thus the Brookhaven experiment gives $ (a_{\mu}^{exp} - a_{\mu}^{theory}) =(287 \pm 80)
\times 10^{-11}$ and if one uses the scaling factor of $m_e^2/m_\mu^2$
the new physics effect in $a_e$  would be of size $(.6\pm .2)\times 10^{-13}$ which is an order of
magnitude smaller than the current error in $\delta \Delta a_e$.
 However, much larger new physics effects can occur if naive scaling law is violated.
 In this work we have shown that such violations do occur  in extensions of MSSM with a
 vectorlike multiplet. In this regard we are in agreement with the conclusion of \cite{Giudice:2012ms}). 
 Thus we have computed the effect of both the non-supersymmetric  as well
 as  the supersymmetric loop corrections to the anomalous magnetic moment of the electron
 in the MSSM extension with a vectorlike multiplet.
 We have shown that effects as large as factors of five or more can occur in the MSSM extension
 over what one expects from scaling. We have also investigated the effect of CP phases
 on the correction from the new physics sector. The largeness of the correction opens
 the possibility that such effects could be discerned even with modest further improvement in
 reducing the error in $\Delta a_e$.

\noindent
{\em Acknowledgments}:
This research was supported in part by the NSF Grant PHY-1314774,
XSEDE- TG-PHY110015, and NERSC-DE-AC02-05CH1123.

\section{Appendix: Further details on the scalar mass  squared matrices   \label{sec9}}
In this Appendix  we give further details of the structure of the slepton mass matrices.
First we discuss briefly the parameters that enter the theory.  The analysis we are doing is
within the framework of an extended MSSM with soft parameters. Thus the analysis is done at the
electroweak scale without renormalization group running. The soft sector of MSSM consists of 
the scalar slepton masses such as $m_0$,  $U(1) (SU(2))$ 
gaugino masses $m_1 (m_2)$ and trilinear couplings such as $A_0^{\tilde \nu}$ where 
$m_1(m_2)$, $A_0^{\tilde \nu}$ are complex with phases $\theta_1, \theta_2, \theta_{A_0^{\tilde \nu}}$ etc.
The other MSSM parameters include $\tan\beta =<H_2>/<H_1>$ where $H_2^2$ gives mass to the 
up quarks while $H_1^1$ gives mass to the down quarks and the leptons, and $\mu$ which is the 
Higgs mixing parameter which can also be complex with the phase $\theta_{\mu}$. The extended
MSSM sector contains the vector lepton masses $m_E$, $m_N$, the mixing parameters 
defined by Eq.(\ref{5}) and soft parameters in the extended sector defined by Eq.(\ref{13}).
 We give now further details.
The mass terms arising from the superpotential  are given by
\beq
{\cal L}^{\rm mass}_F= {\cal L}_C^{\rm mass} +{\cal L}_N^{\rm mass}\ ,
\eeq
where  ${\cal L}_C^{\rm mass}$ gives the mass terms for the charged leptons while
$ {\cal L}_N^{mass}$ gives the mass terms for the  neutrinos. For ${\cal L}_C^{\rm mass}$ we have
\begin{gather}
-{\cal L}_C^{\rm mass} =\left(\frac{v^2_2 |f'_2|^2}{2} +|f_3|^2+|f_3'|^2+|f_3''|^2\right)\tilde E_R \tilde E^*_R
+\left(\frac{v^2_2 |f'_2|^2}{2} +|f_4|^2+|f_4'|^2+|f_4''|^2\right)\tilde E_L \tilde E^*_L\nonumber\\
+\left(\frac{v^2_1 |f_1|^2}{2} +|f_4|^2\right)\tilde \tau_R \tilde \tau^*_R
+\left(\frac{v^2_1 |f_1|^2}{2} +|f_3|^2\right)\tilde \tau_L \tilde \tau^*_L
+\left(\frac{v^2_1 |h_1|^2}{2} +|f_4'|^2\right)\tilde \mu_R \tilde \mu^*_R\nonumber\\
+\left(\frac{v^2_1 |h_1|^2}{2} +|f_3'|^2\right)\tilde \mu_L \tilde \mu^*_L
+\left(\frac{v^2_1 |h_2|^2}{2} +|f_4''|^2\right)\tilde e_R \tilde e^*_R
+\left(\frac{v^2_1 |h_2|^2}{2} +|f_3''|^2\right)\tilde e_L \tilde e^*_L\nonumber\\
+\Bigg\{-\frac{f_1 \mu^* v_2}{\sqrt{2}} \tilde \tau_L \tilde \tau^*_R
-\frac{h_1 \mu^* v_2}{\sqrt{2}} \tilde \mu_L \tilde \mu^*_R
 -\frac{f'_2 \mu^* v_1}{\sqrt{2}} \tilde E_L \tilde E^*_R
+\left(\frac{f'_2 v_2 f^*_3}{\sqrt{2}}  +\frac{f_4 v_1 f^*_1}{\sqrt{2}}\right) \tilde E_L \tilde \tau^*_L\nonumber\\
+\left(\frac{f_4 v_2 f'^*_2}{\sqrt{2}}  +\frac{f_1 v_1 f^*_3}{\sqrt{2}}\right) \tilde E_R \tilde \tau^*_R
+\left(\frac{f'_3 v_2 f'^*_2}{\sqrt{2}}  +\frac{h_1 v_1 f'^*_4}{\sqrt{2}}\right) \tilde E_L \tilde \mu^*_L
+\left(\frac{f'_2 v_2 f'^*_4}{\sqrt{2}}  +\frac{f'_3 v_1 h^*_1}{\sqrt{2}}\right) \tilde E_R \tilde \mu^*_R\nonumber\\
+\left(\frac{f''^*_3 v_2 f'_2}{\sqrt{2}}  +\frac{f''_4 v_1 h^*_2}{\sqrt{2}}\right) \tilde E_L \tilde e^*_L
+\left(\frac{f''_4 v_2 f'^*_2}{\sqrt{2}}  +\frac{f''^*_3 v_1 h^*_2}{\sqrt{2}}\right) \tilde E_R \tilde e^*_R
+f'_3 f^*_3 \tilde \mu_L \tilde \tau^*_L +f_4 f'^*_4 \tilde \mu_R \tilde \tau^*_R\nonumber\\
+f_4 f''^*_4 \tilde {e}_R \tilde{\tau}^*_R
+f''_3 f^*_3 \tilde {e}_L \tilde{\tau}^*_L
+f''_3 f'^*_3 \tilde {e}_L \tilde{\mu}^*_L
+f'_4 f''^*_4 \tilde {e}_R \tilde{\mu}^*_R
-\frac{h_2 \mu^* v_2}{\sqrt{2}} \tilde{e}_L \tilde{e}^*_R
+H.c. \Bigg\}
\end{gather}

For ${\cal L}_N^{\rm mass}$ we have
\begin{multline}
-{\cal L}_N^{\rm mass}=
\left(\frac{v^2_1 |f_2|^2}{2}
 +|f_3|^2+|f_3'|^2+|f_3''|^2\right)\tilde N_R \tilde N^*_R\\
 +\left(\frac{v^2_1 |f_2|^2}{2}+|f_5|^2+|f_5'|^2+|f_5''|^2\right)\tilde N_L \tilde N^*_L
+\left(\frac{v^2_2 |f'_1|^2}{2}+|f_5|^2\right)\tilde \nu_{\tau R} \tilde \nu^*_{\tau R}\\
+\left(\frac{v^2_2 |f'_1|^2}{2}
+|f_3|^2\right)\tilde \nu_{\tau L} \tilde \nu^*_{\tau L}
+\left(\frac{v^2_2 |h'_1|^2}{2}
+|f_3'|^2\right)\tilde \nu_{\mu L} \tilde \nu^*_{\mu L}
+\left(\frac{v^2_2 |h'_1|^2}{2}
+|f_5'|^2\right)\tilde \nu_{\mu R} \tilde \nu^*_{\mu R}\nonumber\\
+\left(\frac{v^2_2 |h'_2|^2}{2}
+|f_3''|^2\right)\tilde \nu_{e L} \tilde \nu^*_{e L}
+\left(\frac{v^2_2 |h'_2|^2}{2}
+|f_5''|^2\right)\tilde \nu_{e R} \tilde \nu^*_{e R}\nonumber\\
+\Bigg\{ -\frac{f_2 \mu^* v_2}{\sqrt{2}} \tilde N_L \tilde N^*_R
-\frac{f'_1 \mu^* v_1}{\sqrt{2}} \tilde \nu_{\tau L} \tilde \nu^*_{\tau R}
-\frac{h'_1 \mu^* v_1}{\sqrt{2}} \tilde \nu_{\mu L} \tilde \nu^*_{\mu R}
+\left(\frac{f_5 v_2 f'^*_1}{\sqrt{2}}  -\frac{f_2 v_1 f^*_3}{\sqrt{2}}\right) \tilde N_L \tilde \nu^*_{\tau L}\nonumber\\
+\left(\frac{f_5 v_1 f^*_2}{\sqrt{2}}  -\frac{f'_1 v_2 f^*_3}{\sqrt{2}}\right) \tilde N_R \tilde \nu^*_{\tau R}
+\left(\frac{h'_1 v_2 f'^*_5}{\sqrt{2}}  -\frac{f'_3 v_1 f^*_2}{\sqrt{2}}\right) \tilde N_L \tilde \nu^*_{\mu L}
+\left(\frac{f''_5 v_1 f^*_2}{\sqrt{2}}  -\frac{f''^*_3 v_2 h'_2}{\sqrt{2}}\right) \tilde N_R \tilde \nu^*_{e R}\nonumber\\
+\left(\frac{h'^*_2 v_2 f''_5}{\sqrt{2}}  -\frac{f''^*_3 v_1 f_2}{\sqrt{2}}\right) \tilde N_L \tilde \nu^*_{e L}
+\left(\frac{f'_5 v_1 f^*_2}{\sqrt{2}}  -\frac{h'_1 v_2 f'^*_3}{\sqrt{2}}\right) \tilde N_R \tilde \nu^*_{\mu R}\nonumber\\
+f'_3 f^*_3 \tilde \nu_{\mu L} \tilde \nu_{\tau^*_L} +f_5 f'^*_5 \tilde \nu_{\mu R} \tilde \nu^*_{\tau R}
-\frac{h'_2 \mu^* v_1}{\sqrt{2}} \tilde{\nu}_{e L} \tilde{\nu}^*_{e R}\\
+f''_3 f^*_3   \tilde{\nu}_{e L} \tilde{\nu}^*_{\tau L}
+f_5 f''^*_5   \tilde{\nu}_{e R} \tilde{\nu}^*_{\tau R}
+f''_3 f'^*_3   \tilde{\nu}_{e L} \tilde{\nu}^*_{\mu L}
+f'_5 f''^*_5   \tilde{\nu}_{e R} \tilde{\nu}^*_{\mu R}
+H.c. \Bigg\}.
\label{11b}
\end{multline}

We define the scalar mass squared   matrix $M^2_{\tilde \tau}$  in the basis $(\tilde  \tau_L, \tilde E_L, \tilde \tau_R,
\tilde E_R, \tilde \mu_L, \tilde \mu_R, \tilde e_L, \tilde e_R)$. We  label the matrix  elements of these as $(M^2_{\tilde \tau})_{ij}= M^2_{ij}$ where the elements of the matrix are given by
\begin{gather}
M^2_{11}=\tilde M^2_{\tau L} +\frac{v^2_1|f_1|^2}{2} +|f_3|^2 -m^2_Z \cos 2 \beta \left(\frac{1}{2}-\sin^2\theta_W\right), \nonumber\\
M^2_{22}=\tilde M^2_E +\frac{v^2_2|f'_2|^2}{2}+|f_4|^2 +|f'_4|^2+|f''_4|^2 +m^2_Z \cos 2 \beta \sin^2\theta_W, \nonumber\\
M^2_{33}=\tilde M^2_{\tau} +\frac{v^2_1|f_1|^2}{2} +|f_4|^2 -m^2_Z \cos 2 \beta \sin^2\theta_W, \nonumber\\
M^2_{44}=\tilde M^2_{\chi} +\frac{v^2_2|f'_2|^2}{2} +|f_3|^2 +|f'_3|^2+|f''_3|^2 +m^2_Z \cos 2 \beta \left(\frac{1}{2}-\sin^2\theta_W\right), \nonumber\\
M^2_{55}=\tilde M^2_{\mu L} +\frac{v^2_1|h_1|^2}{2} +|f'_3|^2 -m^2_Z \cos 2 \beta \left(\frac{1}{2}-\sin^2\theta_W\right), \nonumber\\
M^2_{66}=\tilde M^2_{\mu} +\frac{v^2_1|h_1|^2}{2}+|f'_4|^2 -m^2_Z \cos 2 \beta \sin^2\theta_W, \nonumber
\end{gather}

\begin{gather}
M^2_{77}=\tilde M^2_{e L} +\frac{v^2_1|h_2|^2}{2}+|f''_3|^2 -m^2_Z \cos 2 \beta \left(\frac{1}{2}-\sin^2\theta_W\right), \nonumber\\
M^2_{88}=\tilde M^2_{e} +\frac{v^2_1|h_2|^2}{2}+|f''_4|^2 -m^2_Z \cos 2 \beta \sin^2\theta_W, \nonumber\\
M^2_{12}=M^{2*}_{21}=\frac{ v_2 f'_2f^*_3}{\sqrt{2}} +\frac{ v_1 f_4 f^*_1}{\sqrt{2}} ,\nonumber\\
M^2_{13}=M^{2*}_{31}=\frac{f^*_1}{\sqrt{2}}(v_1 A^*_{\tau} -\mu v_2),\nonumber\\
M^2_{14}=M^{2*}_{41}=0, M^2_{15} =M^{2*}_{51}=f'_3 f^*_3,\nonumber\\
 M^{2*}_{16}= M^{2*}_{61}=0,  M^{2*}_{17}= M^{2*}_{71}=f''_3 f^*_3,  M^{2*}_{18}= M^{2*}_{81}=0,
M^2_{23}=M^{2*}_{32}=0,\nonumber\\
M^2_{24}=M^{2*}_{42}=\frac{f'^*_2}{\sqrt{2}}(v_2 A^*_{E} -\mu v_1),  M^2_{25} = M^{2*}_{52}= \frac{ v_2 f'_3f'^*_2}{\sqrt{2}} +\frac{ v_1 h_1 f^*_4}{\sqrt{2}} ,\nonumber\\
 M^2_{26} =M^{2*}_{62}=0,  M^2_{27} =M^{2*}_{72}=  \frac{ v_2 f''_3f'^*_2}{\sqrt{2}} +\frac{ v_1 h_1 f'^*_4}{\sqrt{2}},  M^2_{28} =M^{2*}_{82}=0, \nonumber\\
M^2_{34}=M^{2*}_{43}= \frac{ v_2 f_4 f'^*_2}{\sqrt{2}} +\frac{ v_1 f_1 f^*_3}{\sqrt{2}}, M^2_{35} =M^{2*}_{53} =0, M^2_{36} =M^{2*}_{63}=f_4 f'^*_4,\nonumber\\
 M^2_{37} =M^{2*}_{73} =0,  M^2_{38} =M^{2*}_{83} =f_4 f''^*_4,
M^2_{45}=M^{2*}_{54}=0, M^2_{46}=M^{2*}_{64}=\frac{ v_2 f'_2 f'^*_4}{\sqrt{2}} +\frac{ v_1 f'_3 h^*_1}{\sqrt{2}}, \nonumber\\
 M^2_{47} =M^{2*}_{74}=0,  M^2_{48} =M^{2*}_{84}=  \frac{ v_2 f'_2f''^*_4}{\sqrt{2}} +\frac{ v_1 f''_3 h^*_2}{\sqrt{2}},\nonumber\\
M^2_{56}=M^{2*}_{65}=\frac{h^*_1}{\sqrt{2}}(v_1 A^*_{\mu} -\mu v_2),
 M^2_{57} =M^{2*}_{75}=f''_3 f'^*_3,  M^2_{58} =M^{2*}_{85}=0,  M^2_{67} =M^{2*}_{76}=0,\nonumber\\
 M^2_{68} =M^{2*}_{86}=f'_4 f''^*_4,  M^2_{78}=M^{2*}_{87}=\frac{h^*_2}{\sqrt{2}}(v_1 A^*_{e} -\mu v_2)
\label{14}
\end{gather}

Here the terms $M^2_{11}, M^2_{13}, M^2_{31}, M^2_{33}$ arise from soft
breaking in the  sector $\tilde \tau_L, \tilde \tau_R$,
the terms $M^2_{55}, M^2_{56}, M^2_{65}, M^2_{66}$ arise from soft
breaking in the  sector $\tilde \mu_L, \tilde \mu_R$,
the terms $M^2_{77}, M^2_{78}, M^2_{87}, M^2_{88}$ arise from soft
breaking in the  sector $\tilde e_L, \tilde e_R$
 and
the terms
$M^2_{22}, M^2_{24},$  $M^2_{42}, M^2_{44}$ arise from soft
breaking in the  sector $\tilde E_L, \tilde E_R$. The other terms arise  from mixing between the staus, smuons and
the mirrors.  We assume that all the masses are of the electroweak size
so all the terms enter in the mass squared  matrix.  We diagonalize this hermitian mass squared  matrix  by the
 unitary transformation
$
 \tilde D^{\tau \dagger} M^2_{\tilde \tau} \tilde D^{\tau} = diag (M^2_{\tilde \tau_1},
M^2_{\tilde \tau_2}, M^2_{\tilde \tau_3},  M^2_{\tilde \tau_4},  M^2_{\tilde \tau_5},  M^2_{\tilde \tau_6},  M^2_{\tilde \tau_7},  M^2_{\tilde \tau_8} )$. {For a further clarification of the notation see~\cite{Ibrahim:2012ds}}.\\

The  mass$^2$  matrix in the sneutrino sector has a similar structure. In the basis $(\tilde  \nu_{\tau L}, \tilde N_L,$
$ \tilde \nu_{\tau R}, \tilde N_R, \tilde  \nu_{\mu L},\tilde \nu_{\mu R}, \tilde \nu_{e L}, \tilde \nu_{e R} )$,  we can write the sneutrino mass$^2$ matrix in the form
$(M^2_{\tilde\nu})_{ij}=m^2_{ij}$ where
\begin{gather}
m^2_{11}=\tilde M^2_{\tau L} +m^2_{\nu_\tau} +|f_3|^2 +\frac{1}{2}m^2_Z \cos 2 \beta,  \nonumber\\
m^2_{22}=\tilde M^2_N +m^2_{N} +|f_5|^2 +|f'_5|^2+|f''_5|^2, \nonumber\\
m^2_{33}=\tilde M^2_{\nu_\tau} +m^2_{\nu_\tau} +|f_5|^2,  \nonumber\\
m^2_{44}=\tilde M^2_{\chi} +m^2_{N} +|f_3|^2 +|f'_3|^2+|f''_3|^2 -\frac{1}{2}m^2_Z \cos 2 \beta, \nonumber\\
m^2_{55}=\tilde M^2_{\mu L} +m^2_{\nu_\mu} +|f'_3|^2 +\frac{1}{2}m^2_Z \cos 2 \beta,  \nonumber\\
m^2_{66}=\tilde M^2_{\nu_\mu} +m^2_{\nu_\mu} +|f'_5|^2,  \nonumber\\
m^2_{77}=\tilde M^2_{e L} +m^2_{\nu_e} +|f''_3|^2+\frac{1}{2}m^2_Z \cos 2 \beta,  \nonumber\\
m^2_{88}=\tilde M^2_{\nu_e} +m^2_{\nu_e} +|f''_5|^2,  \nonumber\\
m^2_{12}=m^{2*}_{21}=\frac{v_2 f_5 f'^*_1}{\sqrt{2}}-\frac{ v_1 f_2 f^*_3}{\sqrt{2}},\nonumber\\
m^2_{13}=m^{2*}_{31}=\frac{f'^*_1}{\sqrt{2}}(v_2 A^*_{\nu_\tau} -\mu v_1),
m^2_{14}=m^{2*}_{41}=0,\nonumber\\
m^2_{15}=m^{2*}_{51}= f'_3 f^*_3, m^2_{16}=m^{2*}_{61}=0,\nonumber\\
m^2_{17}=m^{2*}_{71}= f''_3 f^*_3, m^2_{18}=m^{2*}_{81}=0,\nonumber\\
m^2_{23}=m^{2*}_{32}=0,
m^2_{24}=m^{2*}_{42}=\frac{f^*_2}{\sqrt{2}}(v_{1}A^*_N-\mu v_2), m^2_{25}=m^{2*}_{52}=-\frac{v_{1}f^*_2 f'_3}{\sqrt{2}}+\frac{h'_1 v_2 f'^*_5}{\sqrt{2}},\nonumber\\
m^2_{26}=m^{2*}_{62}=0, m^2_{27}=m^{2*}_{72}=-\frac{v_{1}f^*_2 f''_3}{\sqrt{2}}+\frac{h'_2 v_2 f''^*_5}{\sqrt{2}},
\end{gather}

\begin{gather}
m^2_{28}=m^{2*}_{82}=0, m^2_{34}=m^{2*}_{43}=\frac{v_1 f^*_2 f_5}{\sqrt{2}}-\frac{v_2 f'_1 f^*_3}{\sqrt{2}},\nonumber\\
m^2_{35}=m^{2*}_{53}=0, m^2_{36}=m^{2*}_{63}=f_5 f'^*_5, m^2_{37}=m^{2*}_{73}=0, m^2_{38}=m^{2*}_{83}=f_5 f''^*_5, m^2_{45}=m^{2*}_{54}=0, \nonumber\\
m^2_{46}=m^{2*}_{64}=-\frac{h'^*_1 v_2 f'_3}{\sqrt{2}}+\frac{v_1 f_2 f'^*_5}{\sqrt{2}}, m^2_{47}=m^{2*}_{74}=0, \nonumber\\
m^2_{48}=m^{2*}_{84}=\frac{v_1 f_2 f''^*_5}{\sqrt{2}}-\frac{v_2 h'^*_2 f''_3}{\sqrt{2}}, m^2_{56}=m^{2*}_{65}=\frac{h'^*_1}{\sqrt{2}}(v_2 A^*_{\nu_\mu}-\mu v_1), \nonumber\\
m^2_{57}=m^{2*}_{75}= f''_3 f'^*_3, m^2_{58}=m^{2*}_{85}=0, m^2_{67}=m^{2*}_{76}=0, \nonumber\\
m^2_{68}=m^{2*}_{86}= f'_5 f''^*_5, m^2_{78}=m^{2*}_{87}=\frac{h'^*_2}{\sqrt{2}}(v_2 A^*_{\nu_e}-\mu v_1).  
\label{15}
\end{gather}

As in the charged  slepton sector here also the terms $m^2_{11}, m^2_{13}, m^2_{31}, m^2_{33}$ arise from soft breaking in the  sector $\tilde \nu_{\tau L}, \tilde \nu_{\tau R}$, the terms $m^2_{55}, m^2_{56}, m^2_{65}, m^2_{66}$ arise from soft breaking in the  sector $\tilde \nu_{\mu L}, \tilde \nu_{\mu_R}$,
the terms $m^2_{77}, m^2_{78}, m^2_{87}, m^2_{88}$ arise from soft
breaking in the  sector $\tilde \nu_{eL}, \tilde \nu_{eR}$
 and
the terms $m^2_{22}, m^2_{24},$  $m^2_{42}, m^2_{44}$ arise from soft breaking in the  sector $\tilde N_L, \tilde N_R$. The other terms arise from mixing between the physical sector and the mirror sector.  Again as in the charged lepton sector we assume that all the masses are of the electroweak size so all the terms enter in the mass$^2$ matrix.  This mass$^2$  matrix can be diagonalized  by the unitary transformation $ \tilde D^{\nu\dagger} M^2_{\tilde \nu} \tilde D^{\nu} = \text{diag} (M^2_{\tilde \nu_1}, M^2_{\tilde \nu_2}, M^2_{\tilde \nu_3},  M^2_{\tilde \nu_4},M^2_{\tilde \nu_5},  M^2_{\tilde \nu_6}, M^2_{\tilde \nu_7}, M^2_{\tilde \nu_8} )$.

\end{document}